\documentclass[aps,amsmath,nofootinbib,superscriptaddress,amssymb,preprintnumbers,floatfix,showpacs,preprint]{revtex4-1}
 \usepackage{amsmath,txfonts,longtable,booktabs,overpic,amssymb,bm,bbm,multirow,float,graphicx,color,dcolumn,subfigure,hyperref,tikz}
\definecolor{blue}{RGB}{45,48,146}
\hypersetup{colorlinks,citecolor=black,anchorcolor=black,menucolor=black, linkcolor=black,filecolor=black,runcolor=black,urlcolor=black,frenchlinks=true}

\begin{document}

\title{Spin resolved momentum spectra for vacuum pair production via a generalized two level model}

\author{Orkash Amat}
\affiliation{Key Laboratory of Beam Technology of the Ministry of Education, and School of Physics and Astronomy, Beijing Normal University, Beijing 100875, China}
\affiliation{School of Astronomy and Space Science, Nanjing University, Nanjing 210023, China}

\author{Hong-Hao Fan}
\affiliation{Key Laboratory of Beam Technology of the Ministry of Education, and School of Physics and Astronomy, Beijing Normal University, Beijing 100875, China}

\author{Suo Tang}
\affiliation{College of Physics and Optoelectronic Engineering, Ocean University of China, Qingdao, Shandong, 266100, China}

\author{Yong-Feng Huang}
\affiliation{School of Astronomy and Space Science, Nanjing University, Nanjing 210023, China}
\affiliation{Key Laboratory of Modern Astronomy and Astrophysics (Nanjing University), Ministry of Education, Nanjing 210023, China}

\author{Bai-Song Xie}\email{Correspondence Author. email: bsxie@bnu.edu.cn}
\affiliation{Key Laboratory of Beam Technology of the Ministry of Education, and School of Physics and Astronomy, Beijing Normal University, Beijing 100875, China}
\affiliation{Institute of Radiation Technology, Beijing Academy of Science and Technology, Beijing 100875, China}

\begin{abstract}
We have formulated a generalized two level model for studying the pair production in multidimensional time-dependent electric fields. It can provide momentum spectra with fully spin resolved components for all possible combined spin states of the particle and anti-particle simultaneously. Moreover, we have also investigated the validity of the two level model for fermions (scalar particles) by comparing the results with those by equal-time Dirac-Heisenberg-Wigner (Feshbach-Villars-Heisenberg-Wigner) formalism in different regimes of pair creation, i.e., multiphoton and tunneling dominated mechanisms. It is found that the results are consistent with each other, indicating the good approximation of the two level model. In particular, in terms of the two level model, we found that the contribution of the particle momentum spectra is the greatest when the spin states of the particle and anti-particle are parallel with $S=1$. It is believed that by this two level model one can extend researches on pair production for more different background fields, such as a slowly varying spatial-temporal one. Many other interesting phenomena may also be revealed, including the spin-resolved vortex structure that is contained in the phase feature of the distribution function of the created pairs.
\end{abstract}
\maketitle

\section{Introduction}\label{sec:1}

Accompanying the development of quantum electrodynamics (QED), the electron and positron pair creation from QED vacuum under strong background field has been always an interesting research topic \cite{Dirac:1928hu,Klein:1929zz,Sauter:1931zz,Heisenberg:1936nmg,Schwinger:1951nm,
Heinzl:2009bmy,Cartlidge:2018,Abramowicz:2019gvx,Ghenescu:2022ttk,Gonoskov:2021hwf,Fedotov:2022ely,Ranken:2024gpg,Xie:2017}. We know that the Schwinger critical field strength is so high as $E_{cr}=m^{2}c^{3}/e\hbar \thickapprox 1.3 \times 10^{16}\rm{V/cm}$ (the corresponding laser intensity is about $4.3 \times 10^{29}\rm{W/cm^2}$, where $m$ and $-e$ are the electron mass and charge) that the current optical (photon wavelength and energy are in $\mu{\rm m}$ and ${\rm eV}$ level) or X-ray (photon wavelength and energy are in ${\rm nm}$ and ${\rm keV}$ level) laser intensity is incapable to reach or/and approach it yet \cite{ELI,XCELS,Abramowicz:2021zja}. Upon the theoretical aspects, the already operating X-ray free electron laser (XFEL) facilities both at Linac Coherent Light Source (LCLS) in the Stanford Linear Accelerator Center (SLAC) and the DESY can in principle achieve near-critical field strength as large as $E\thickapprox 0.1 E_{cr}$, corresponding to a laser intensity of about $7\times 10^{27}\rm{W/cm^2}$ \cite{Ringwald:2001ib}. The future experiment has been estimated numerically \cite{Alkofer:2001ik}.
Unfortunately, direct observation of the vacuum pair production still awaits experimental verification.

On the other hand, it is possible to observe the vacuum pair production when the field strength is less than the critical field $E_{cr}$.
For instance, the vacuum pair production can be studied through multiphoton,
tunneling and multi-mechanism (including multiphoton and tunneling processes) regimes when the field strength is smaller than the critical field $E_{cr}$ \cite{Keldysh:1964ud,Keldysh:1965ojf}.
The vacuum pair production could be investigated via different methods, for example,
the worldline instanton (WI) technique \cite{Affleck:1981bma,Dunne:2005sx,Dunne:2006st,Dunne:2006ur,Dunne:2006ff,Dunne:2008zza,Schutzhold:2008pz,Dunne:2008kc,Dumlu:2011cc,BaisongXie:2012,Ilderton:2015lsa,Ilderton:2015qda,Schneider:2014mla,Schneider:2018huk,Rajeev:2021zae,DegliEsposti:2021its,DegliEsposti:2022yqw,Amat:2022uxq,Horvathy:2023vpy,Copinger:2023ctz,Drummond:2023wqc,Manzo:2024gto},
the real time Dirac-Heisenberg-Wigner (DHW) formalism~\cite{Bialynicki-Birula:1991jwl,Hebenstreit:2011wk,Hebenstreit:2011,Kohlfurst:2015zxi,Olugh:2018seh,Ababekri:2019dkl,Kohlfurst:2019mag,Aleksandrov:2019ddt,Li:2021wag,Kohlfurst:2021dfk,Kohlfurst:2021skr,Mohamedsedik:2021pzb,Li:2021vjf,Hu:2022ouk,Kohlfurst:2022edl,Kohlfurst:2022vwf,Hu:2023pmz,Amat:2023vwv,Chen:2024gkx},
the computational quantum field theory ~\cite{Krekora:2004trv,Krekora:2004,Tang:2013qna,Wang:2019oyk,Wang:2021tmo,Li:2023neo,Gong:2024fon,Sawut:2024sze},
the imaginary time method ~\cite{Popov:2005rp}, the quantum Vlasov equation (QVE) approach~\cite{Kluger:1998bm,Schmidt:1998vi,Hebenstreit:2009km,Li:2014xga,Li:2014psw,Gong:2020jqs},
the Wentzel-Kramers-Brillouin (WKB) approach~\cite{Popov:1971iga,Kim:2007pm,Dumlu:2010ua,Dumlu:2011rr,Strobel:2014tha,Dumlu:2015paa,Oertel:2016vsg,Dumlu:2022olg,Lysenko:2023wrs},
scattering matrix approach ~\cite{Ritus:1985,Titov:2018bgy,Tang:2019ffe,Ilderton:2019ceq,King:2019igt,Seipt:2020diz,Wistisen:2020rsq,Lei:2021eqe,Podszus:2021lms,Bu:2021ebc,Tang:2022tmn,Golub:2022cvd,MacLeod:2022qid,Gao:2022cfe,Tang:2023cyz,Lu:2023wrf,Zhao:2023nbl,Ababekri:2022mob,ababekri2024generation},
quantum two level model (TLM) \cite{Piazza:2001,Avetissian:2002ucr,Akkermans:2011yn,Melike_2012,Kaminski:2018ywj,Krajewska:2019vqd,Fiordilino:2021zkp,Dunne:2022zlx,Bechler:2023kjx}, and so on.
Here we want to stress that the DHW formalism could theoretically solve the vacuum pair production problem for any background field~\cite{Kohlfurst:2019mag},
providing spin-dependent positron / electron momentum spectrum \cite{Blinne:2015zpa,Blinne:2016yzv}. Recently, the spin effect on the pair production under temporal electric fields has been investigated by using the DHW formalism \cite{Kohlfurst:2018kxg,Hu:2024nyp}, focusing on how the spin effect influences the momentum spiral and asymmetry degree in the vacuum pair production. In addition, spin resolved momentum spectra for vacuum pair production has been investigated by using scattering matrix approach \cite{Majczak:2024hmt,Aleksandrov:2024rsz}, which is applicable to study the vacuum pair production in arbitrary time-dependent electric fields. To our knowledge, the momentum spectrum depends on the spin states of the electron and positron simultaneously, which however cannot be obtained by this method. In this regard, therefore, we make an effort to derive a generalized TLM that can provide full spin resolved momentum spectra for any spin state of created pair. Meanwhile, up to now, the TLM in vacuum pair production has only been applicable to linearly polarized electric fields. Therefore, it is necessary to develop a generalized TLM that can be used to investigate vacuum pair production in multidimensional time-dependent electric fields.

In this paper, we formulate a generalized fermionic TLM that can provide full spin resolved momentum spectra for any spin state of created pair simultaneously. The method is applicable to study the vacuum pair production in multidimensional time-dependent electric fields. Moreover, we also investigate the validity of present formulated TLM for fermions by comparing the results with those by the DHW formalism in different pair creation regimes, i.e., multiphoton and tunneling dominated mechanisms.

The paper is organized as follows.
In Sec.~\ref{sec:2}, we  introduce the generalized fully spin resolved fermionic TLM under arbitrarily time-dependent electric fields.
In Sec.~\ref{sec:3}, we representatively demonstrate an external background field.
In Sec.~\ref{sec:4}, we show the numerical results of the generalized fully spin resolved fermionic TLM and DHW for various regimes, as well as study the spin effect and validity of them.
The summary is given in Sec.~\ref{sec:5}. For convenience,  scalar TLM, equal-time Feshbach-Villars-Heisenberg-Wigner (FVHW) \cite{Feshbach:1958wv,Best:1992gb,Best:1993wq,Zhuang:1995pd,Zhuang:1998bqx} and DHW formalisms are briefly derived in the Appendices.

We use the natural units ($\hbar=c=1$) throughout this paper, and express all quantities in terms of the electron mass $m$.
\section{Theoretical framework}\label{sec:2}

In order to study the fermionic vacuum pair creation under more realistic strong electric field,
we need to obtain a more general TLM form.
In this section, we introduce generalized fully spin resolved fermionic TLM in an arbitrarily time-dependent external background field. We consider the fermionic pair production in an arbitrary time-dependent electric field $\mathbf{E}(t)=(E_x(t), E_y(t), E_z(t))$, which corresponds a vector potential $\mathbf{A}(t)=(A_x(t), A_y(t), A_z(t))$.
The relationship between the electric field and vector potential is $\mathbf{E}(t)=-\dot{\mathbf{A}}(t)$.

The first step is to solve the Dirac equation under the external field
\begin{equation}\label{B3}
\left( i \gamma^\mu \partial_{\mu} -  e \gamma^\mu A_\mu  - m \right) \Psi(\mathbf{x},t) = 0,
\end{equation}
where the Dirac gamma matrices $\gamma^{\mu}$ are
\begin{equation}\label{B4}
\gamma^0=\left(
           \begin{array}{cc}
            \mathbbm{1}_2 & 0 \\
             0 & -\mathbbm{1}_2 \\
           \end{array}
         \right), \quad
\bm{\gamma}=\left(
           \begin{array}{cc}
             0 & \bm{\sigma} \\
             -\bm{\sigma} & 0 \\
           \end{array}
         \right), \quad
\end{equation}
and the Pauli matrices are
\begin{align}\label{B5}
  \sigma_1=\left(
    \begin{array}{cc}
      0 & 1 \\
      1 & 0 \\
    \end{array}
  \right), \sigma_2=\left(
    \begin{array}{cc}
      0 & -i \\
      i & 0 \\
    \end{array}
  \right), \sigma_3=\left(
    \begin{array}{cc}
      1 & 0 \\
      0 & -1 \\
    \end{array}
  \right).
\end{align}
For the arbitrary time-dependent electric field $A^\mu(t)=(0,~\mathbf{A}(t))$, the wave function can be decomposed as \cite{Marinov:1977gq}
\begin{eqnarray}\label{B6}
\Psi(\mathbf{x},t)=\int \frac{d^3\mathbf{q}}{(2 \pi)^3}  e^{i\mathbf{q}\cdot \mathbf{x}} \psi_{\mathbf{q}}(t).
\end{eqnarray}
One can get a Schr\"{o}dinger like equation after substituting Eq.~\eqref{B6} into Eq.~\eqref{B3}
\begin{subequations}
\begin{align}\label{B7a}
\hat{H}_\mathbf{q}(t) \psi_{\mathbf{q}}(t)&=i\partial_{t} \psi_{\mathbf{q}}(t),\\ \label{B7b}
H_\mathbf{q}(t) \psi_{\mathbf{q}}(t)&= \left[ \gamma^0 \bm{\gamma} \cdot\left(\mathbf{q} - e \mathbf{A}(t)\right) + \gamma^0 m \right] \psi_{\mathbf{q}}(t).
\end{align}
\end{subequations}
We can introduce the kinetic momentum $\mathbf{p}(t)=\mathbf{q} - e \mathbf{A}(t)$, where $\mathbf{q}$ repesents the canonical momentum. The Hamiltonian can be written as
\begin{align}\label{B8}
H_\mathbf{q}(t)=
\left(
\begin{array}{cccc}
       m  & 0 &  p_z (t) &  (p_x (t)-ip_y (t)) \\
       0 & m  &  (p_x (t)+ip_y (t)) & - p_z (t) \\
       p_z (t) &  (p_x (t)-ip_y (t)) & -m & 0 \\
      (p_x (t)+ip_y (t)) & - p_z (t) & 0 & -m  \\
\end{array}
\right),
\end{align}
where $H_{\mathbf{q}}^2(t) = m^2 + {[\mathbf{q} - e \mathbf{A}(t)]}^2 = \omega_{\mathbf{q}}^2(t)$. The self-adjoint operator $\hat{H}_{\mathbf{q}}(t) = \hat{H}_{\mathbf{q}}^\dagger(t)$ has the instantaneous eigenvectors
\begin{align}\label{B9}
\hat{H}_{\bf{q}}(t) u_{\bf{q},\bf{s}}(t) &= H_{\bf{q}}(t) u_{\bf{q},\bf{s}}(t)=+ \omega_{\mathbf{q}}(t) u_{\bf{q},\bf{s}}(t),\\ \label{B10}
\hat{H}_{\bf{q}}(t) v_{\bf{q},\bf{s}}(t) &= H_{\bf{q}}(t) v_{\bf{q},\bf{s}}(t)=- \omega_{\mathbf{q}}(t) v_{\bf{q},\bf{s}}(t),
\end{align}
in which $u^\dagger_{\bf{q},\bf{s}}(t) u_{\bf{q},\bf{s}'}(t) = v^\dagger_{\bf{q},\bf{s}}(t) v_{\bf{q},\bf{s}'}(t)=  \delta_{\bf{s} \bf{s}'}$ and $u^\dagger_{\bf{q},\bf{s}}(t) v_{\bf{q},\bf{s}'}(t)= v^\dagger_{\bf{q},\bf{s}}(t) u_{\bf{q},\bf{s}'}(t)= 0$.

We can obtain the following expression after taking the derivative of Eqs.~\eqref{B9} and ~\eqref{B10} over time and multiplying $u^\dagger_{\bf{q},\bf{s}'}(t)$ and $v^\dagger_{\bf{q},\bf{s}'}(t)$
\begin{align}\label{B11}
\dot{u}^\dagger_{\bf{q},\bf{s}'}(t) v_{\bf{q},\bf{s}}(t) &= \biggl( v^\dagger_{\bf{q},\bf{s}'}(t) \dot{u}_{\bf{q},\bf{s}}(t)\biggr)^{*} = +\frac{1}{2 \omega_{\bf q}(t)} \biggl(u^\dagger_{\bf{q},\bf{s}'}(t) \dot{H}_{\bf{q}}(t) v_{\bf{q},\bf{s}}(t)\biggr),\\ \label{B12}
\dot{v}^\dagger_{\bf{q},\bf{s}'}(t) u_{\bf{q},\bf{s}}(t) &= \biggl( u^\dagger_{\bf{q},\bf{s}'}(t) \dot{v}_{\bf{q},\bf{s}}(t)\biggr)^{*} = -\frac{1}{2 \omega_{\bf q}(t)} \biggl(v^\dagger_{\bf{q},\bf{s}'}(t) \dot{H}_{\bf{q}}(t) u_{\bf{q},\bf{s}}(t)\biggr).
\end{align}
$\dot{u}^\dagger_{\bf{q},\bf{s}'}(t) u_{\bf{q},\bf{s}}(t)$ and $\dot{v}^\dagger_{\bf{q},\bf{s}'}(t) v_{\bf{q},\bf{s}}(t)$ terms can be neglected in slowly varying external fields. We will show the details and plausibility in Sec.~\ref{sec:4}.

The time-dependent eigenvectors of the Hamiltonian Eq.~\eqref{B8} can be obtained in the light cone \cite{Ritus:1985vta}
\begin{equation}\label{B15}
\begin{split}
  u_{\mathbf{q},+1}(t)&=\frac{1}{\sqrt{4\omega_\mathbf{q}(t)(\omega_\mathbf{q}(t)- p_z(t))}}\left(
     \begin{array}{c}
       \omega_\mathbf{q}(t)+m - p_z(t) \\
       -(p_x(t)+ip_y(t)) \\
       -\omega_\mathbf{q}(t)+m + p_z(t) \\
       (p_x(t)+ip_y(t)) \\
     \end{array}
   \right)\!,\;\;\\
   u_{\mathbf{q},-1}(t)&=\!\frac{1}{\sqrt{4\omega_\mathbf{q}(t)(\omega_\mathbf{q}(t)- p_z(t))}}\left(
     \begin{array}{c}
       (p_x(t)-ip_y(t)) \\
       \omega_\mathbf{q}(t)+m - p_z(t) \\
       (p_x(t)-ip_y(t)) \\
       \omega_\mathbf{q}(t)-m - p_z(t) \\
     \end{array}
   \right)\!,\, \\
v_{\mathbf{q},+1}(t)&=\!\frac{1}{\sqrt{4\omega_\mathbf{q}(t)(\omega_\mathbf{q}(t)+ p_z(t))}}\left(
    \begin{array}{c}
      (p_x(t)-ip_y(t)) \\
      -\omega_\mathbf{q}(t)+m - p_z(t) \\
      (p_x(t)-ip_y(t)) \\
      -\omega_\mathbf{q}(t)-m - p_z(t) \\
    \end{array}
  \right)\!,\, \\
v_{\mathbf{q},-1}(t)&=\frac{1}{\sqrt{4\omega_\mathbf{q}(t)(\omega_\mathbf{q}(t)+ p_z(t))}}\left(
     \begin{array}{c}
       -\omega_\mathbf{q}(t)+m - p_z(t) \\
       -(p_x(t)+ip_y(t)) \\
       \omega_\mathbf{q}(t)+m + p_z(t) \\
       (p_x(t)+ip_y(t)) \\
     \end{array}
   \right),
\end{split}
\end{equation}
where $s$ denotes the spin of the particle, and positive ($+1$) and negative ($-1$) values correspond to the spin states of up and down, respectively.

We can expand $\psi_{\mathbf{q}}(t)$ as
\begin{equation}\label{B16}
\psi_{\mathbf{q}}(t) = \alpha_{\mathbf{q}}(t)  u_{\bf{q},\bf{s}}(t) e^{-i \int^t d\tau \omega_{\mathbf{q}}(\tau)} +\beta_{\mathbf{q}}(t)  v_{\bf{q},\bf{s}}(t) e^{i \int^t d\tau \omega_{\mathbf{q}}(\tau)}.
\end{equation}
Additionally, one can demonstrate by using Eqs.~\eqref{B7a}, ~\eqref{B7b}, ~\eqref{B9}, ~\eqref{B10} and ~\eqref{B16} that
\begin{align}\label{B17}
\begin{split}
\hat{H}_\mathbf{q}(t) \psi_{\mathbf{q}}(t)=&i \partial_{t} \psi_{\mathbf{q}}(t)=i \biggl(\dot{\alpha}_{\mathbf{q}}(t)  u_{\bf{q},\bf{s}}(t) e^{-i \int^t d\tau \omega_{\mathbf{q}}(\tau)} + \alpha_{\mathbf{q}}(t)  \dot{u}_{\bf{q},\bf{s}}(t) e^{-i \int^t d\tau \omega_{\mathbf{q}}(\tau)}\\
& + (-i \omega_{\mathbf{q}}) \alpha_{\mathbf{q}}(t)  u_{\bf{q},\bf{s}}(t) e^{-i \int^t d\tau \omega_{\mathbf{q}}(\tau)}+\dot{\beta}_{\mathbf{q}}(t)  v_{\bf{q},\bf{s}}(t) e^{i \int^t d\tau \omega_{\mathbf{q}}(\tau)} \\
&+\beta_{\mathbf{q}}(t)  \dot{v}_{\bf{q},\bf{s}}(t) e^{i \int^t d\tau \omega_{\mathbf{q}}(\tau)} +  (i \omega_{\mathbf{q}})   \beta_{\mathbf{q}}(t)  v_{\bf{q},\bf{s}}(t) e^{i \int^t d\tau \omega_{\mathbf{q}}(\tau)}  \biggr),
\end{split}\\ \label{B18}
H_\mathbf{q}(t) \psi_{\mathbf{q}}(t)=&\omega_{\mathbf{q}} \alpha_{\mathbf{q}}(t)  u_{\bf{q},\bf{s}}(t) e^{-i \int^t d\tau \omega_{\mathbf{q}}(\tau)} - \omega_{\mathbf{q}} \beta_{\mathbf{q}}(t)  v_{\bf{q},\bf{s}}(t) e^{i \int^t d\tau \omega_{\mathbf{q}}(\tau)}.
\end{align}
After multiplying $u^\dagger_{\bf{q},\bf{s}'}(t)$ and $v^\dagger_{\bf{q},\bf{s}'}(t)$ again by the left and right sides of Eqs.~\eqref{B17} and ~\eqref{B18}, we can further prove that
\begin{align}\label{B19}
\begin{split}
  \omega_{\mathbf{q}} \alpha_{\mathbf{q}}(t) e^{-i \int^t d\tau \omega_{\mathbf{q}}(\tau)}=&i \biggl(\dot{\alpha}_{\mathbf{q}}(t) e^{-i \int^t d\tau \omega_{\mathbf{q}}(\tau)} + \alpha_{\mathbf{q}}(t) u^\dagger_{\bf{q},\bf{s}'}(t) \dot{u}_{\bf{q},\bf{s}}(t) e^{-i \int^t d\tau \omega_{\mathbf{q}}(\tau)}\\
  & + (-i \omega_{\mathbf{q}}) \alpha_{\mathbf{q}}(t)  e^{-i \int^t d\tau \omega_{\mathbf{q}}(\tau)}+\beta_{\mathbf{q}}(t) u^\dagger_{\bf{q},\bf{s}'}(t)  \dot{v}_{\bf{q},\bf{s}}(t) e^{i \int^t d\tau \omega_{\mathbf{q}}(\tau)}\biggr),
\end{split}\\ \label{B20}
\begin{split}
  \omega_{\mathbf{q}} \beta_{\mathbf{q}}(t) e^{i \int^t d\tau \omega_{\mathbf{q}}(\tau)}=&-i \biggl(\dot{\beta}_{\mathbf{q}}(t) e^{i \int^t d\tau \omega_{\mathbf{q}}(\tau)} + \alpha_{\mathbf{q}}(t) v^\dagger_{\bf{q},\bf{s}'}(t) \dot{u}_{\bf{q},\bf{s}}(t) e^{-i \int^t d\tau \omega_{\mathbf{q}}(\tau)}\\
  & + (i \omega_{\mathbf{q}}) \beta_{\mathbf{q}}(t)  e^{i \int^t d\tau \omega_{\mathbf{q}}(\tau)}+\beta_{\mathbf{q}}(t) v^\dagger_{\bf{q},\bf{s}'}(t)  \dot{v}_{\bf{q},\bf{s}}(t) e^{i \int^t d\tau \omega_{\mathbf{q}}(\tau)}\biggr).
  \end{split}
\end{align}
We can get a simple form by substituting Eqs.~\eqref{B11} and \eqref{B12} into Eqs.~\eqref{B19} and \eqref{B20}
\begin{align}\label{B21}
\begin{split}
\omega_{\mathbf{q}} \alpha_{\mathbf{q}}(t) e^{-i \int^t d\tau \omega_{\mathbf{q}}(\tau)}=&i \biggl(\dot{\alpha}_{\mathbf{q}}(t) e^{-i \int^t d\tau \omega_{\mathbf{q}}(\tau)} -  \frac{\beta_{\mathbf{q}}(t)}{2 \omega_{\bf{q}}} \biggl(u^\dagger_{\bf{q},\bf{s}'}(t) \dot{H}_{\bf{q}}(t) v_{\bf{q},\bf{s}}(t)\biggr) e^{i \int^t d\tau \omega_{\mathbf{q}}(\tau)} \\
& + (-i \omega_{\mathbf{q}}) \alpha_{\mathbf{q}}(t)  e^{-i \int^t d\tau \omega_{\mathbf{q}}(\tau)}\biggr),
\end{split}\\ \label{B22}
\begin{split}
\omega_{\mathbf{q}} \beta_{\mathbf{q}}(t) e^{i \int^t d\tau \omega_{\mathbf{q}}(\tau)}=&-i \biggl(\dot{\beta}_{\mathbf{q}}(t) e^{i \int^t d\tau \omega_{\mathbf{q}}(\tau)} +  \frac{\alpha_{\mathbf{q}}(t)}{2 \omega_{\bf{q}}} \biggl(u^\dagger_{\bf{q},\bf{s}'}(t) \dot{H}_{\bf{q}}(t) v_{\bf{q},\bf{s}}(t)\biggr)^* e^{-i \int^t d\tau \omega_{\mathbf{q}}(\tau)}\\
& + (i \omega_{\mathbf{q}}) \beta_{\mathbf{q}}(t)  e^{i \int^t d\tau \omega_{\mathbf{q}}(\tau)}\biggr).
\end{split}
\end{align}

One can obtain two coupled ordinary differential equation as
\begin{align} \label{B24}
\dot{\alpha}_{\bf{q},\bf{s},\bf{s}'}(t) &= \frac{\beta_{\bf{q},\bf{s},\bf{s}'}(t)}{2 \omega_{\mathbf{q}}(t)} e^{2i \int^t d\tau \omega_{\mathbf{q}}(\tau)} \biggl(u^\dagger_{\bf{q},\bf{s}'}(t) \dot{H}_{\bf{q}}(t) v_{\bf{q},\bf{s}}(t)\biggr), \\ \label{B25}
\dot{\beta}_{\bf{q},\bf{s},\bf{s}'}(t) &= -\frac{\alpha_{\bf{q},\bf{s},\bf{s}'}(t)}{2 \omega_{\mathbf{q}}(t)} e^{-2i \int^t d\tau \omega_{\mathbf{q}}(\tau)} {\biggl( u^\dagger_{\bf{q},\bf{s}'}(t) \dot{H}_{\bf{q}}(t) v_{\bf{q},\bf{s}}(t) \biggr)}^*.
\end{align}
This spin-dependent Bogoliubov transformation implements a change from the time-independent creation $a_{\bf{q},\bf{s},\bf{s}'}$ and annihilation operators $b_{-\bf{q},\bf{s},\bf{s}'}^\dagger$ to a time-dependent creation $\tilde{a}_{\bf{q},\bf{s},\bf{s}'}(t)$ and annihilation operators $\tilde{b}_{-\bf{q},\bf{s},\bf{s}'}^\dagger(t)$:
\begin{eqnarray}\label{B25}
\begin{pmatrix}
\tilde{a}_{\bf{q},\bf{s},\bf{s}'}(t)\cr
\tilde{b}_{-\bf{q},\bf{s},\bf{s}'}^\dagger(t)
\end{pmatrix}
=\begin{pmatrix}
\alpha_{\bf{q},\bf{s},\bf{s}'} & -\beta_{\bf{q},\bf{s},\bf{s}'}^*\cr
\beta_{\bf{q},\bf{s},\bf{s}'} & \alpha_{\bf{q},\bf{s},\bf{s}'}^*
\end{pmatrix}
\begin{pmatrix}
a_{\bf{q},\bf{s},\bf{s}'}\cr
b^\dagger_{-\bf{q},\bf{s},\bf{s}'}
\end{pmatrix},
\end{eqnarray}
where $|\alpha_{\bf{q},\bf{s},\bf{s}'}(t)|^2+|\beta_{\bf{q},\bf{s},\bf{s}'}(t)|^2=1$ for the spinor QED.

Now, we redefine the useful coefficients \cite{Akkermans:2011yn}
\begin{eqnarray}\label{B26}
c^{(1)}_{\bf{q},\bf{s},\bf{s}'}(t)&=&\alpha_{\bf{q},\bf{s},\bf{s}'}(t)e^{-i\int^t d\tau \omega_{\mathbf{q}}(\tau)},\\ \label{B27}
c^{(2)}_{\bf{q},\bf{s},\bf{s}'}(t)&=&\beta_{\bf{q},\bf{s},\bf{s}'}(t)e^{i\int^t d\tau \omega_{\mathbf{q}}(\tau)}.
\end{eqnarray}
By the same way, after taking the derivatives of time on the above equations and organizing them, we can obtain the spin-dependent fermionic TLM
\begin{equation}\label{B28}
i\frac{d}{dt}\begin{bmatrix}
c_{\bf{q},\bf{s},\bf{s}'}^{(1)}(t)\\
c_{\bf{q},\bf{s},\bf{s}'}^{(2)}(t)
\end{bmatrix}
=
\begin{pmatrix} \omega_{\bf{q}}(t) & i\Omega_{\bf{q},\bf{s},\bf{s}'}(t) \cr -i\Omega^*_{\bf{q},\bf{s},\bf{s}'}(t) & -\omega_{\bf{q}}(t) \end{pmatrix}
\begin{bmatrix}
c_{\bf{q},\bf{s},\bf{s}'}^{(1)}(t)\\
c_{\bf{q},\bf{s},\bf{s}'}^{(2)}(t)
\end{bmatrix},
\end{equation}
where
\begin{equation}\label{B29}
\Omega_{\bf{q},\bf{s},\bf{s}'}(t)=\frac{u^\dagger_{\bf{q},\bf{s}'}(t) \dot{H}_{\bf{q}}(t) v_{\bf{q},\bf{s}}(t)}{2\omega_{\bf{q}}(t)},
\end{equation}
and the initial conditions are $c_{\bf{q},\bf{s},\bf{s}'}^{(1)}(t_0)=1$ and $c_{\bf{q},\bf{s},\bf{s}'}^{(2)}(t_0)=0$.

The momentum distribution for different spin of electron and positron can be calculated by using the coefficient $c_{\bf{q},\bf{s},\bf{s}'}^{(2)}$ at $t=+\infty$ as
\begin{eqnarray}\label{B30}
f_{\bf{q}}^{\bf{s}\bf{s}'}=2|c_{\bf{q},\bf{s},\bf{s}'}^{(2)}(t=+\infty)|^2.
\end{eqnarray}
The electron and positron  momentum distributions are
\begin{equation}
f_{\bf{q}}^{\bf{s}'}=\sum_{\bf s} f_{\bf{q}}^{\bf{s}\bf{s}'},\ \ f_{\bf{q}}^{\bf{s}}=\sum_{\bf s'} f_{\bf{q}}^{\bf{s}\bf{s}'}.
\end{equation}
The total momentum distribution is
\begin{eqnarray}\label{B31}
f_{\bf{q}}=\sum_{\bf s} \sum_{\bf s'} f_{\bf{q}}^{\bf{s}\bf{s}'}.
\end{eqnarray}
\section{External field configuration}\label{sec:3}
To study the validity of the fermionic TLM, we representatively choose a circularly polarized (CP) electric field, which is obtained from two counter-propagating short laser pulses. Note that the magnetic fields of the two laser pulses can cancel in the standing-wave, and the electric field is approximately spatially homogeneous in all interaction region. The electric field can be written in a time-dependent form
\begin{align}\label{eq:1}
{\bf E}\left(\varphi\right)=\varepsilon_{0} E_{cr} d(\varphi) \begin{pmatrix} {\rm cos}\varphi \cr {\rm sin}\varphi \cr 0 \end{pmatrix},
\end{align}
with an envelope
\begin{align}\label{eq:29}
d(\varphi) & =\begin{cases}
\text{sin}^{4}\left(\frac{\varphi}{2N}\right), & 0<\varphi<2\pi N,\\
0, & \text{otherwise},
\end{cases}
\end{align}
where $\varepsilon_{0}$ is the peak field strength, $E_{cr}$ denotes the Schwinger critical field strength, $\varphi=\omega_{0} t$ is the time-dependent phase of background field, $\omega_{0}$ and $N$ are the frequency and cycle number of the individual electric field, respectively. The pulse duration $\tau$ can be written as $\tau=2\pi N/\omega_0$. We would like to stress that a slowly varying envelope approximation can be adopted to the vacuum pair production for $N \gg 1$.
\section{Fermionic pair production}\label{sec:4}
In this section we study the validity of the generalized fully spin resolved fermionic TLM and compare it with the DHW method (see Appendix \ref{ap:d}). We also investigate the spin effect of the electron and positron on the pair production.

In order to comprehensively test the consistency of these two methods, we will discuss the cases of multiphoton ($\gamma_{\omega} \gg 1$) and tunneling ($\gamma_{\omega} \ll 1$) processes separately, where the Keldysh parameters are defined as $\gamma_{\omega} = \frac{m\omega_{0}}{e\varepsilon_{0}E_{cr}}$ \cite{Blinne:2016yzv,Kohlfurst:2019mag}.

\subsection{Multiphoton dominated process}\label{sec:4A}
\begin{figure}[ht!]\centering
\includegraphics[width=0.49\textwidth]{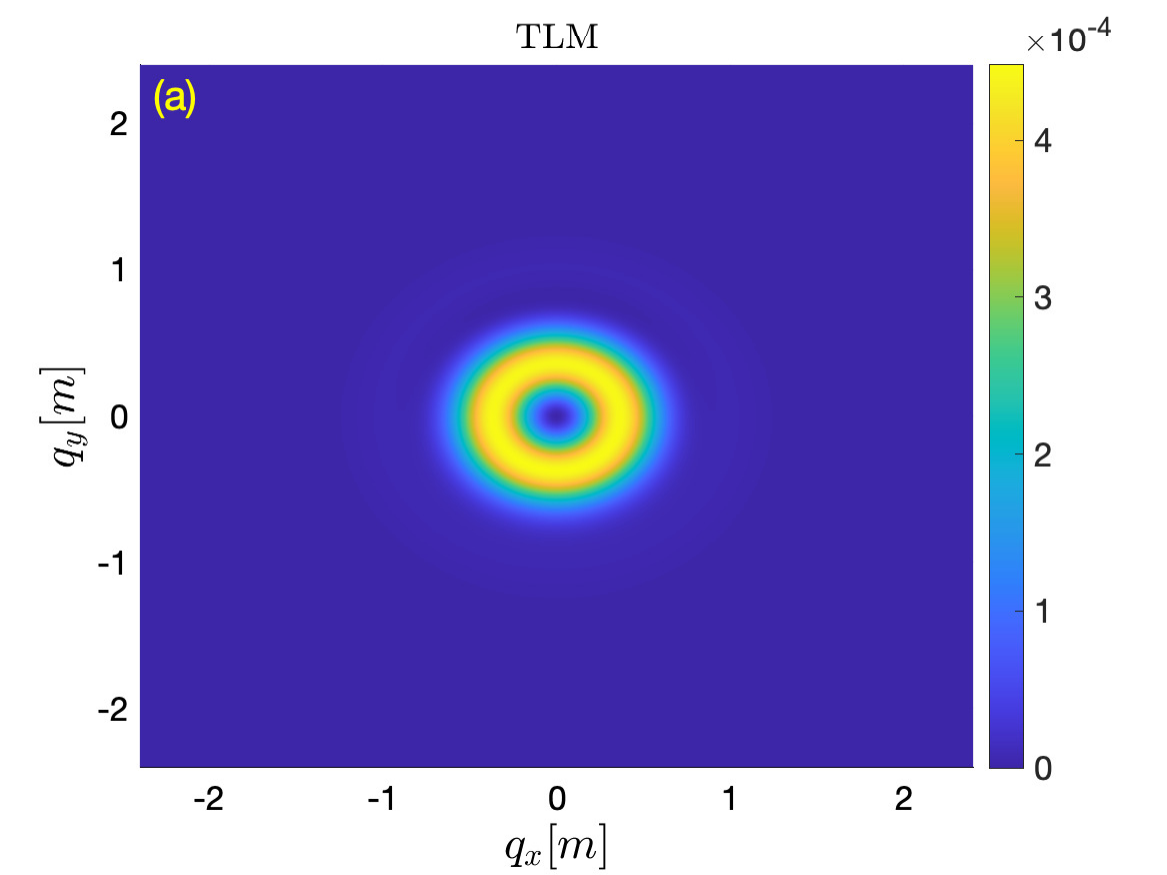}
\includegraphics[width=0.49\textwidth]{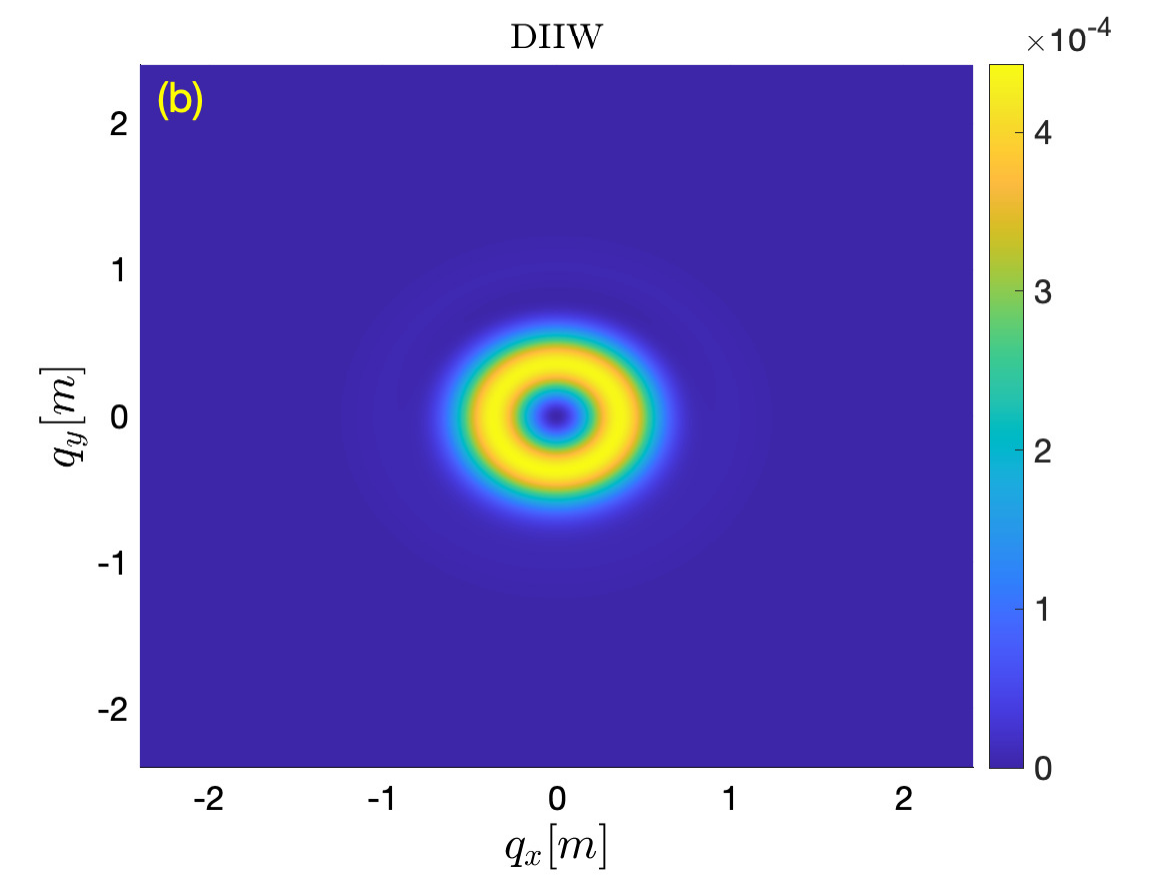}
\caption{Total electron momentum spectra under the CP electric field (where $q_z=0$). (a) and (b) are the numerical results of fermionic TLM and DHW methods. Other field parameters are $\varepsilon_{0}=0.1$, $N=5$ and $\omega_{0}=0.99 m$.
\label{fig:1}}
\end{figure}
The virtual electron in the vacuum can absorb photons from the background field and obtain energy to produce real electron.
Therefore, the energy of an electron can be determined by the number of photons absorbed and the photon energy. For fermionic pair creation, we will use a background field with the field strength of $\varepsilon_{0} = 0.1$, field frequency of $\omega_0 =  0.99 m$ and number of cycles $N$ = 5, respectively. Nothe that we may use the slowly varying envelope approximation since $N>1$.
The Keldysh parameter is $\gamma_{\omega} = 9.9$, indicating that the multiphoton process dominates \cite{Keldysh:1965ojf}.

The total electron momentum spectra of the fermionic TLM and DHW methods for the multiphoton dominated process are shown in Figs.~\ref{fig:1}(a) and (b).
One can mainly find one ring in the electron momentum spectra for both the fermionic TLM and DHW methods.
The pulse duration is about $\tau\thickapprox 32 m^{-1}$.
If we assume that it is a large pulse, the effective mass may be written as $m_{*}\thickapprox m\sqrt{1+\varepsilon_0^2 m^2/4\omega_0^2}$.
Consequently, the photon number could be estimated by $n=2\sqrt{m_{*}^2+q_{n}^2}/\omega_{0}$, i.e., $q_2\thickapprox 0.47m$ for $n=2$ photons. Therefore, the ring structure comes from multiphoton dominated process by absorbing $n=2$ photons.
Most interestingly, we find that the two results obtained from the fermionic TLM and DHW methods are substantially the same, i.e.,
the ratio of the two momentum spectra $R_{\rm F}=f_{\rm TLM}/f_{\rm DHW}$ is about $1$.
This shows that the two methods are consistent in the multiphoton dominated process.
Moreover, the electron momentum spectrum is mainly related to the total energy of electron $\omega_{\mathbf{q}}(t)=\sqrt{m^2 + [q_x - e A_{x}(t)]^2+ [q_y - e A_{y}(t)]^2}$, in which the vector potentials $A_{x}(t)$ and $A_{y}(t)$ are odd and even functions with respect to $t$. Therefore, $\omega_{\mathbf{q}}(t)$ stays invariant when replacing $t \rightarrow -t$ and $q_x \rightarrow -q_x$. This means that all momentum spectra in our work are exactly symmetric with respect to $q_{x}=0$. On the other side, $q^{max}_{2,y}\gg A^{max}_{y}(t)$ so that $\omega_{\mathbf{q}}(t)\thickapprox \sqrt{m^2 + [q_x - e A_{x}(t)]^2+ q_y^2}$ stays invariant. Hence, the momentum spectra for multiphoton dominated process are approximately symmetric with respect to $q_{y}=0$.

\begin{figure}[ht!]\centering
\includegraphics[width=0.49\textwidth]{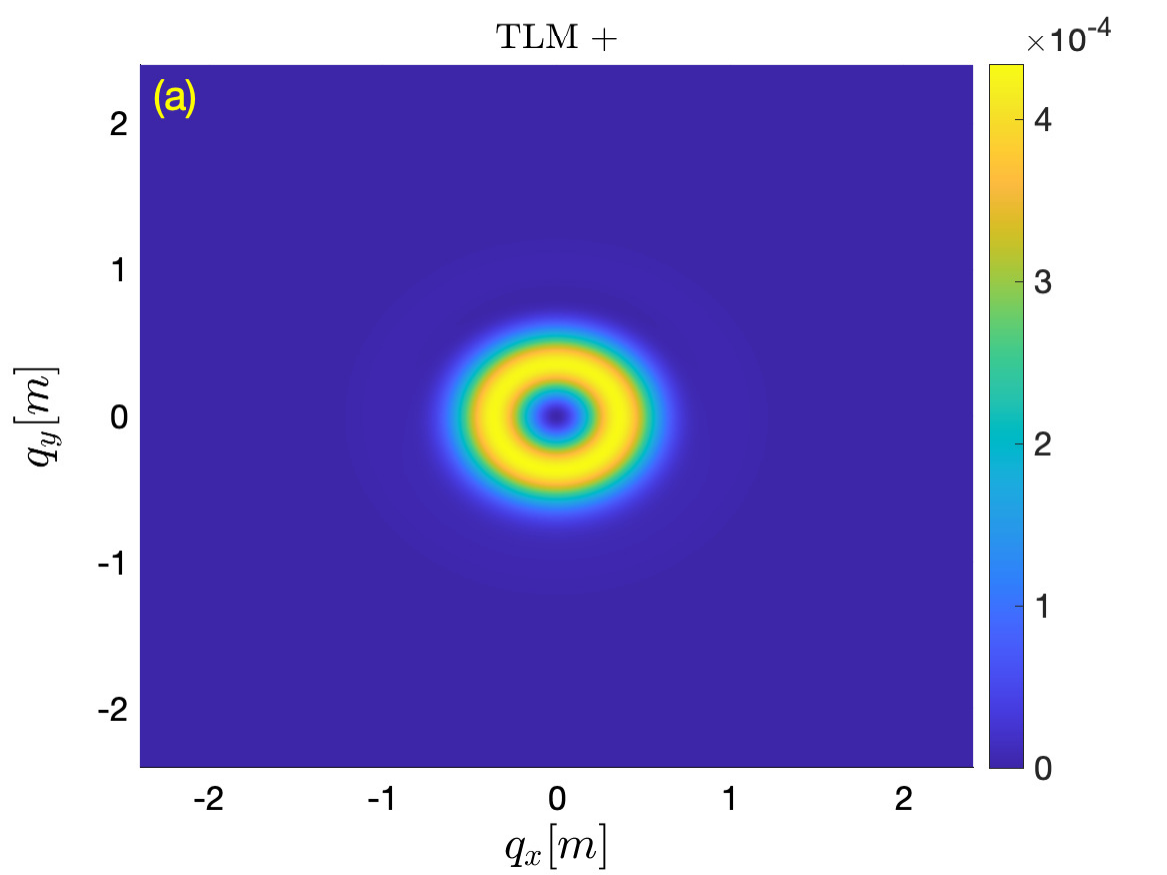}
\includegraphics[width=0.49\textwidth]{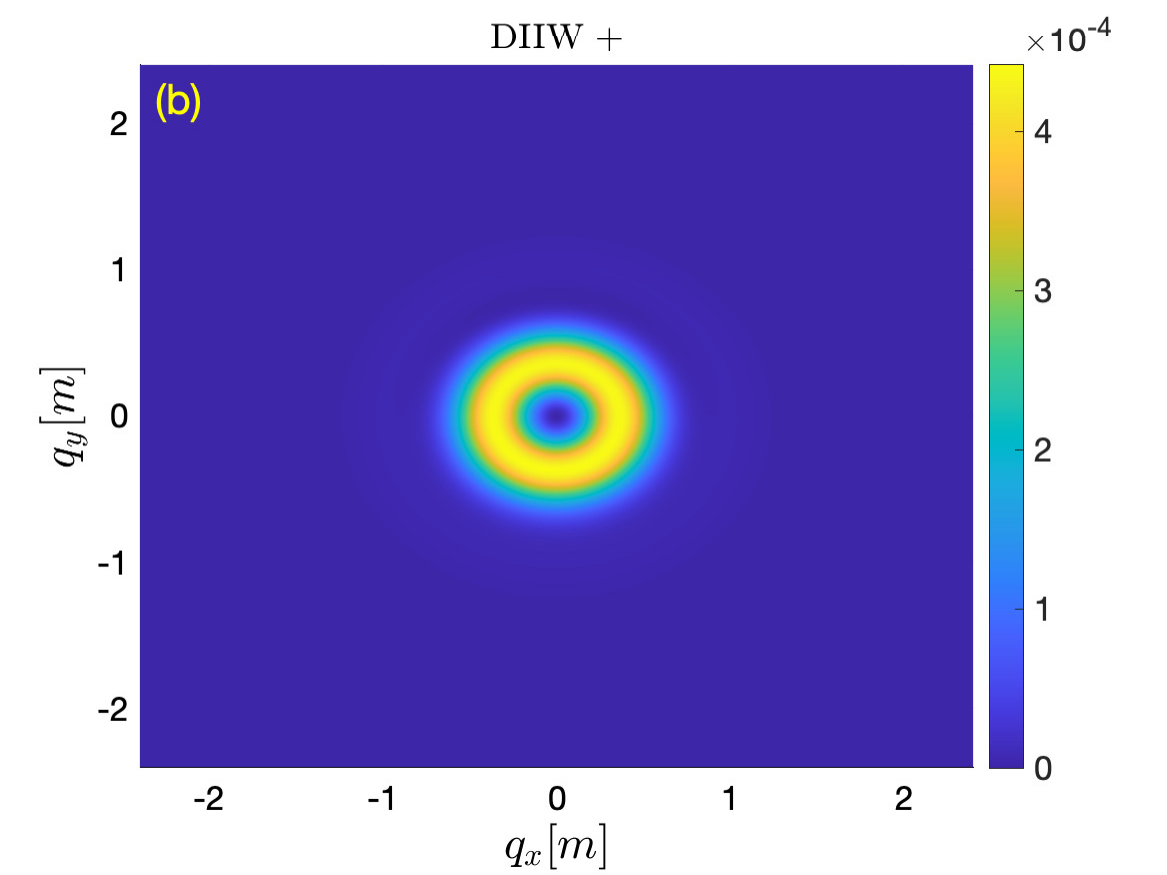}
\includegraphics[width=0.49\textwidth]{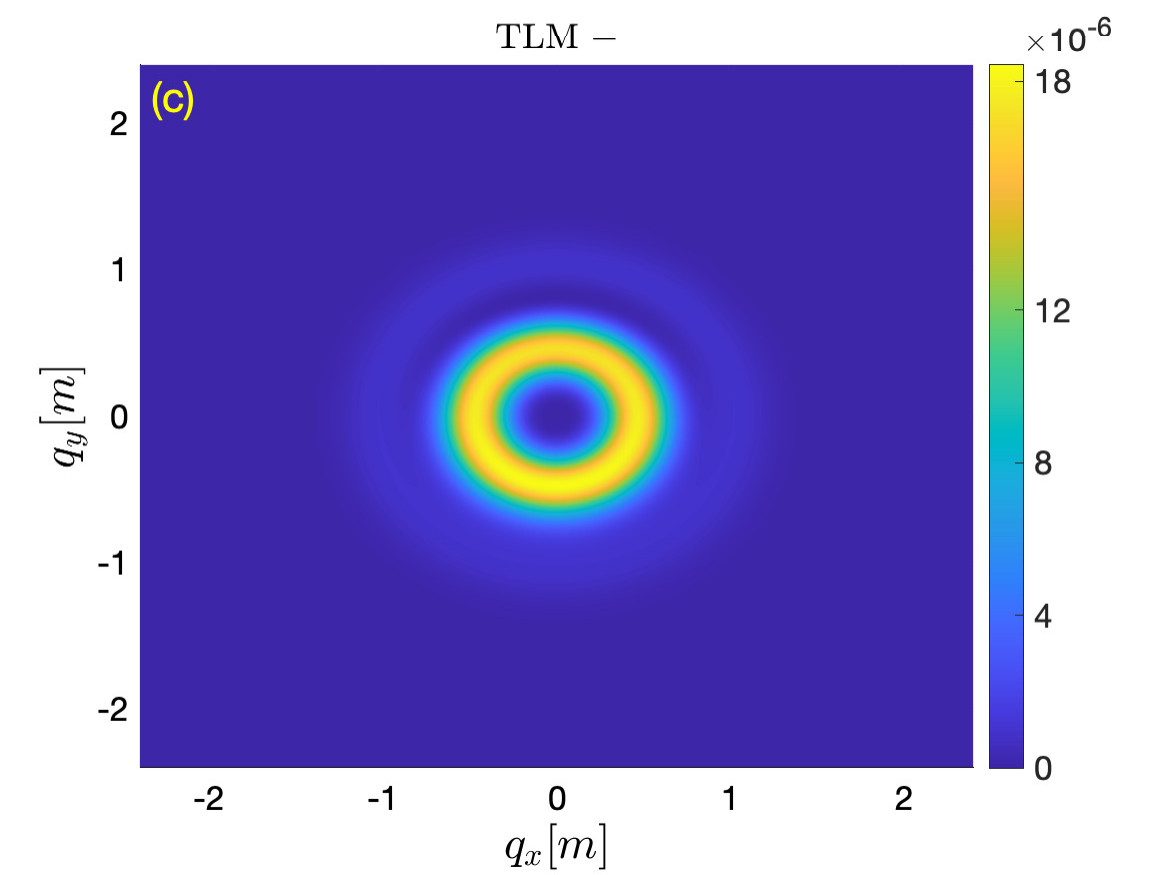}
\includegraphics[width=0.49\textwidth]{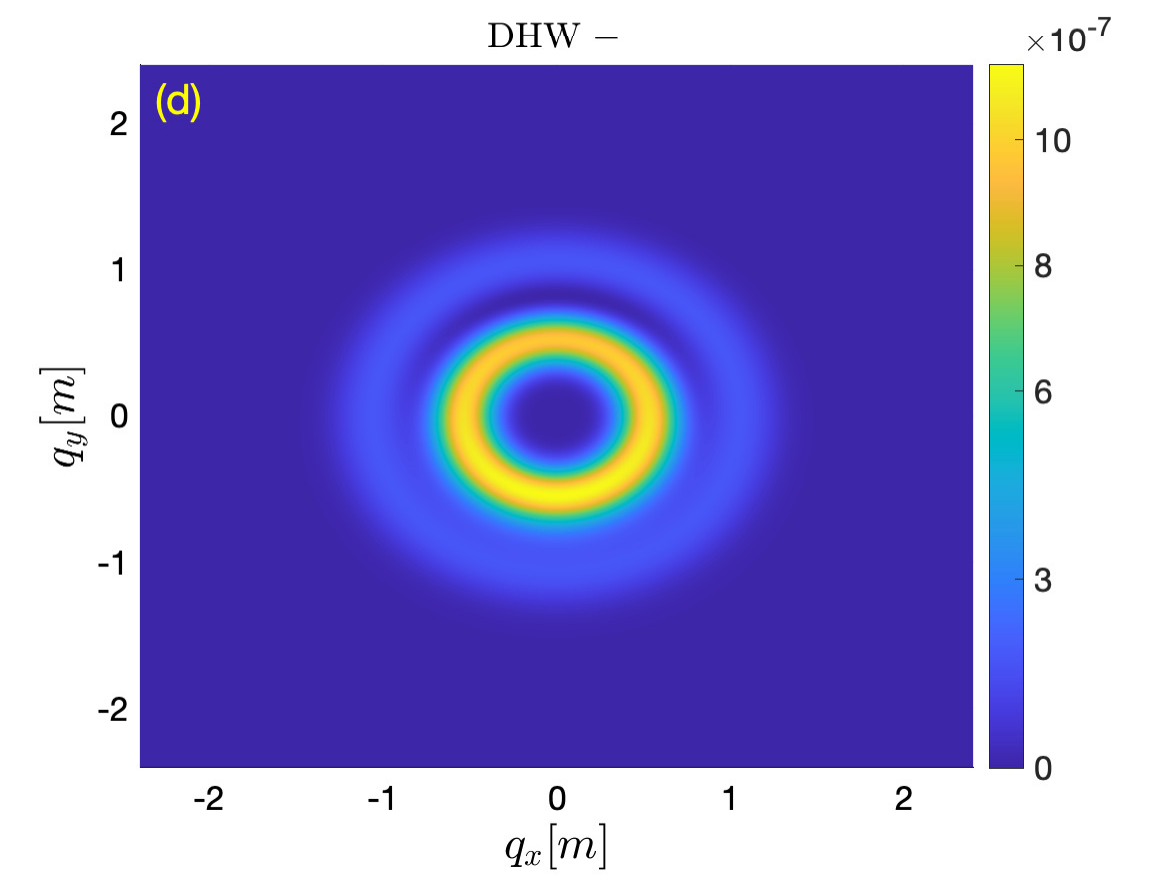}
\caption{Electron momentum spectra under the CP electric field with different method and spin states (where $q_z=0$). The first and second columns represent the fermionic TLM and DHW results respectively. The first and second rows represent the results of electron spin up ($+$) and down ($-$), respectively. Other field parameters are $\varepsilon_{0}=0.1$, $N=5$ and $\omega_{0}=0.99 m$.
\label{fig:2}}
\end{figure}

The spin-dependent momentum spectra for the fermionic TLM and DHW methods are shown in Figs.~\ref{fig:2}(a), (b), (c) and (d).
We further find that the spin-dependent momentum spectra for the fermionic TLM and DHW methods are relatively similar in terms of the general trend, but there are slight differences.
For instance, two rings on the spectra can be obviously seen on the spectra $f^-_{\rm TLM}$and $f^-_{\rm DHW}$ in Figs.~\ref{fig:2}(c) and (d).
Moreover, one can obtain the $R^+_F=f^+_{\rm TLM}/f^+_{\rm DHW}\thickapprox 0.97$ and $R^-_F=f^-_{\rm TLM}/f^-_{\rm DHW}\thickapprox 1.63$.
It could be found that the main contribution of the total electron momentum spectra in Figs.~\ref{fig:1}(a) and (b) comes from spin $+$ electron momentum spectra in Figs.~\ref{fig:2}(a) and (b).
Meanwhile, the two rings in Figs.~\ref{fig:2}(c) and (d) can not be seen in Figs.~\ref{fig:1}(a) and (b).
This explains why we can see one ring in Figs.~\ref{fig:1}(a) and (b).
Note that the sum of the spin $+$ and $-$ electron momentum spectra of the fermionic TLM or DHW methods is equivalent to the electron momentum spectra of the fermionic TLM or DHW methods.

\begin{figure}[ht!]\centering
\includegraphics[width=0.49\textwidth]{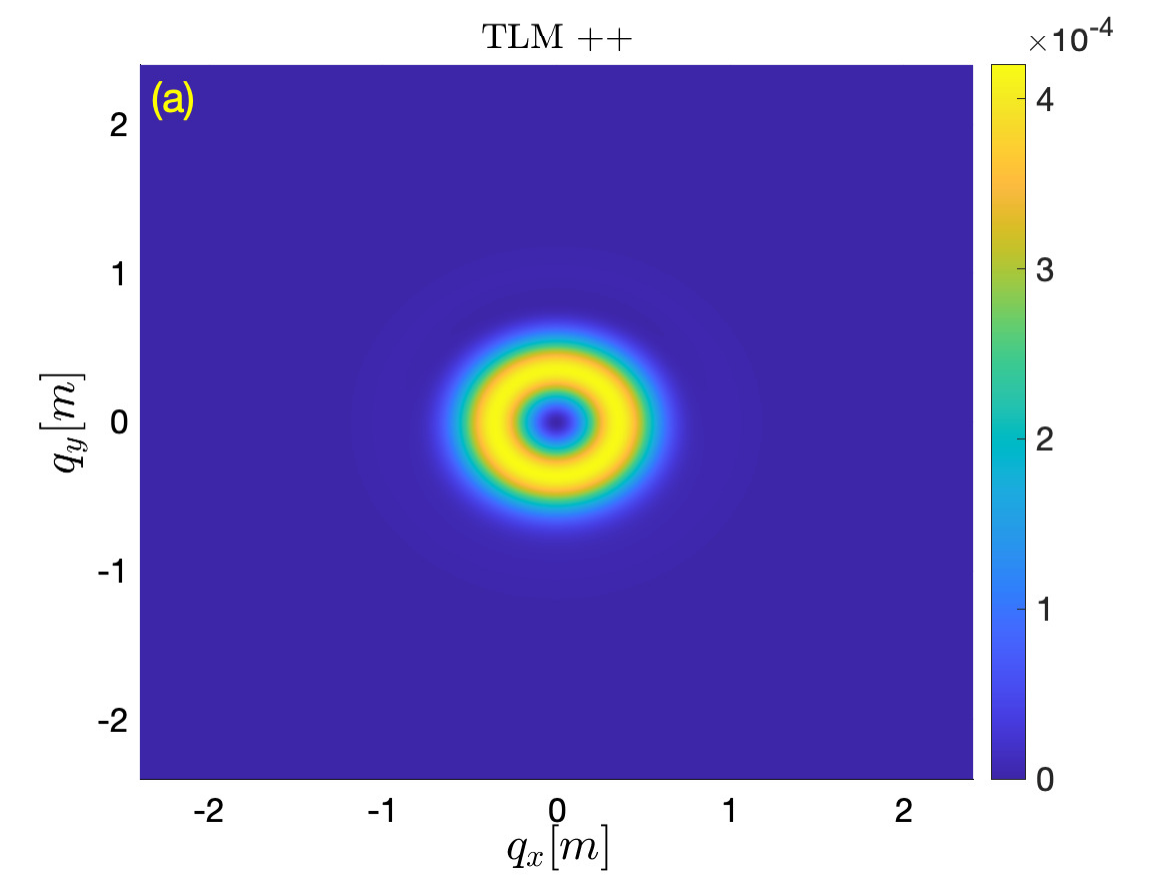}
\includegraphics[width=0.49\textwidth]{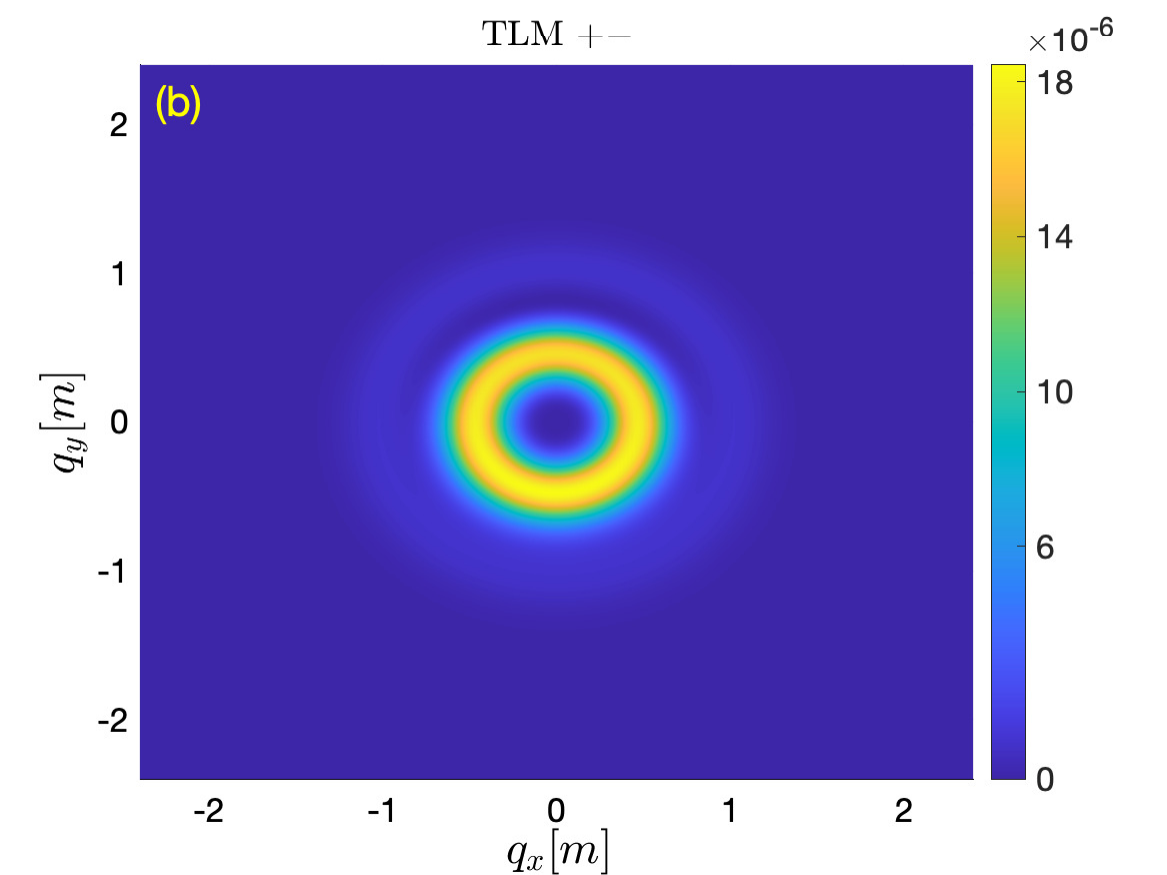}
\includegraphics[width=0.49\textwidth]{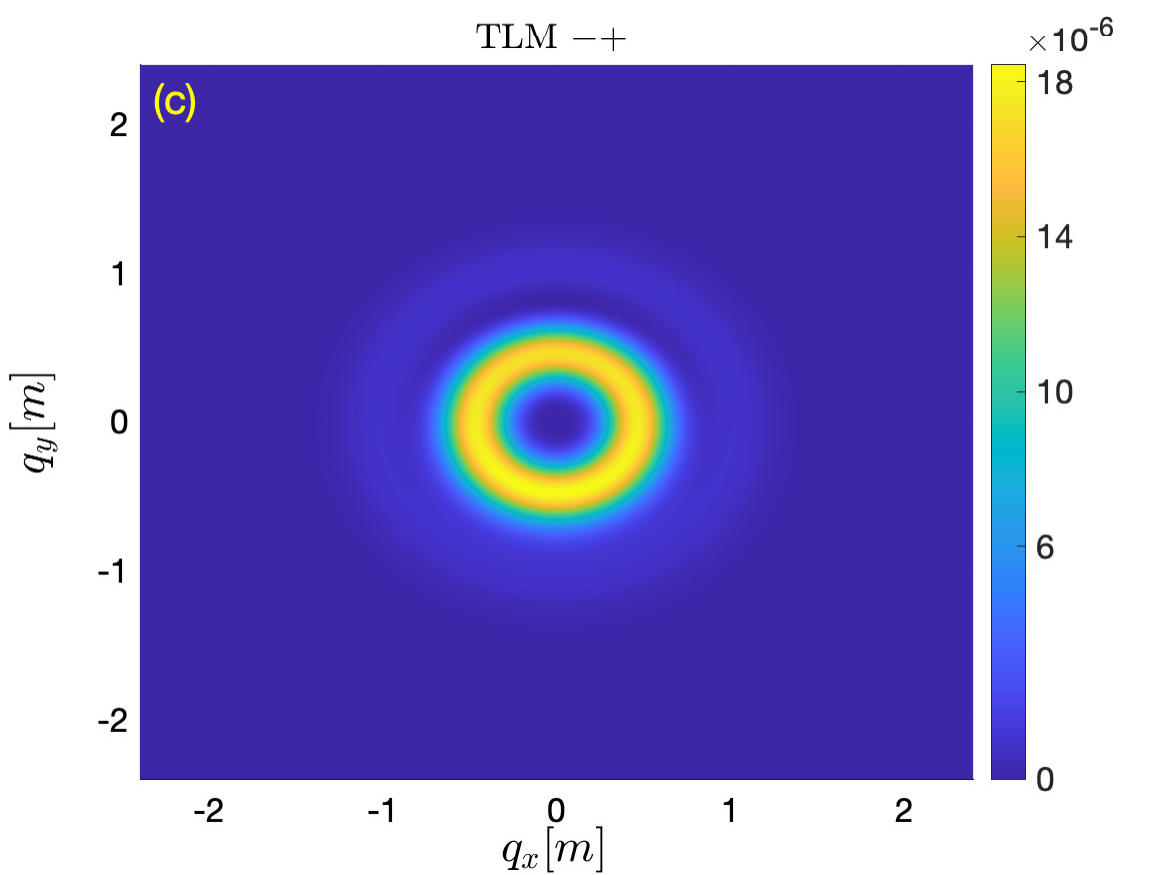}
\includegraphics[width=0.49\textwidth]{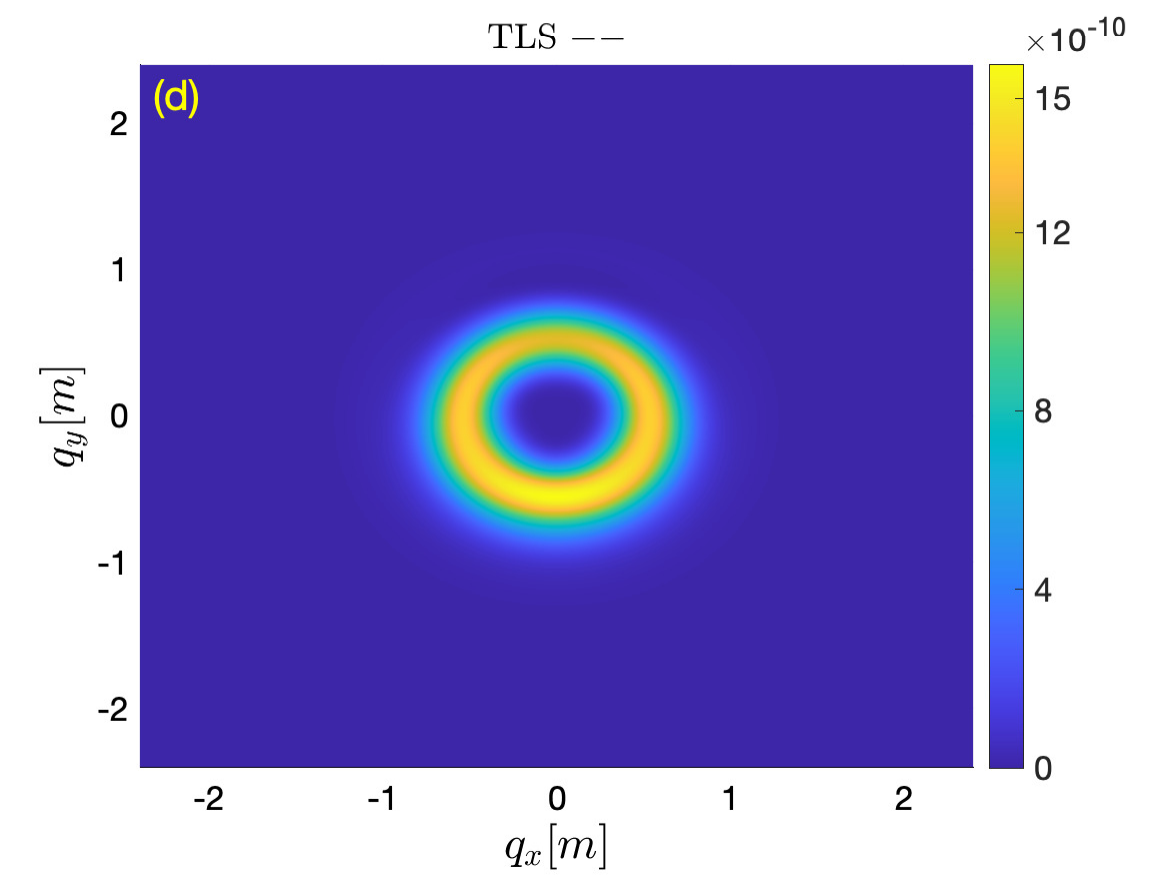}
\caption{Momentum spectra $f^{{\bf s}_{e^+}{\bf s}_{e^-}}_{\rm TLM}$ with different electron and positron spin states via fermionic TLM (where $q_z=0$). The first and second rows represent positron spin $+$ and $-$, respectively. The first and second columns represent the results of electron spin $+$ and $-$, respectively. Other field parameters are $\varepsilon_{0}=0.1$, $N=5$ and $\omega_{0}=0.99 m$.
\label{fig:3}}
\end{figure}

The purpose of introducing the fermionic TLM is that it can explain the electron and positron spin effect on vacuum pair production within external background field.
The full spin resolved momentum spectra $f^{{\bf s}_{e^+}{\bf s}_{e^-}}_{\rm TLM}$ for the fermionic TLM are shown in Figs.~\ref{fig:3}(a), (b), (c) and (d).
One can interpret the results in Figs.~\ref{fig:2}(a), (b), (c) and (d).
For example, the $f^{+}_{\rm TLM}$ in Fig.~\ref{fig:2}(a) equals to $f^{++}_{\rm TLM}+f^{-+}_{\rm TLM}$.
Hence, one ring can be seen in the electron momentum spectrum, as shown in Fig.~\ref{fig:2}(a), because $f^{++}_{\rm TLM}/f^{-+}_{\rm TLM} \thickapprox  20$.
By the same way, one can find that $f^{-}_{\rm TLM}$ in Fig.~\ref{fig:2}(c) is equivalent to sum up all positron spin dependent momentum spectra $f^{+-}_{\rm TLM}+f^{--}_{\rm TLM}$.
It can be seen from Fig.~\ref{fig:2}(c) that the two rings can be observed in the electron momentum spectrum since $f^{+-}_{\rm TLM}/f^{--}_{\rm TLM} \thickapprox  10^4$.
Most notably, we found that the momentum spectrum $f^{++}_{\rm TLM}$ is the greatest among all the spin alignment of the electron and positron. For fermionic vacuum pair production, there are two cases for the spin alignment. If the spin directions of the electron and positron are in the same direction, the total spin is given by $S=1$ for parallel; otherwise $S=0$ for anti-parallel. We obviously find that our momentum spectrum for $S = 1$ is larger than that for $S = 0$, which is consistent with the result in Ref. \cite{Kohlfurst:2018kxg}.

\subsection{Tunneling dominated process}\label{sec:4B}

\begin{figure}[ht!]\centering
\includegraphics[width=0.49\textwidth]{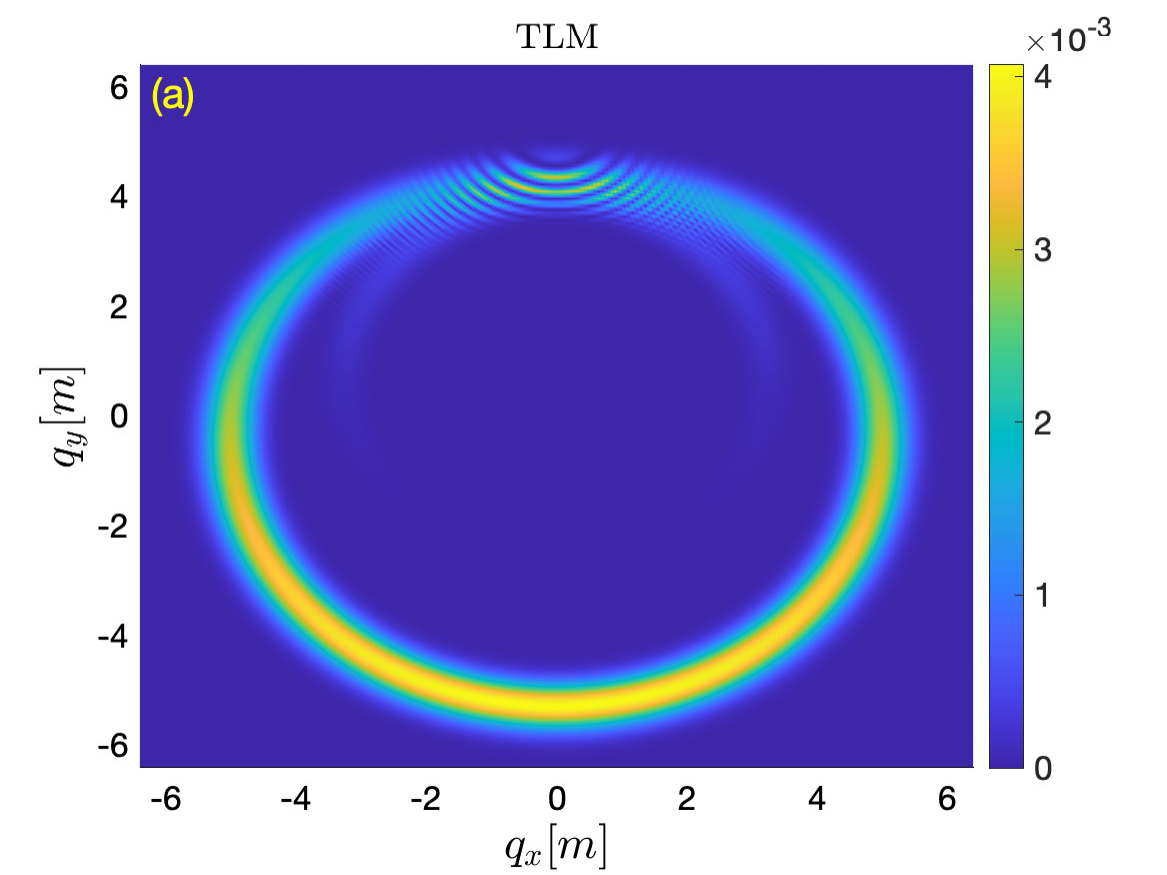}
\includegraphics[width=0.49\textwidth]{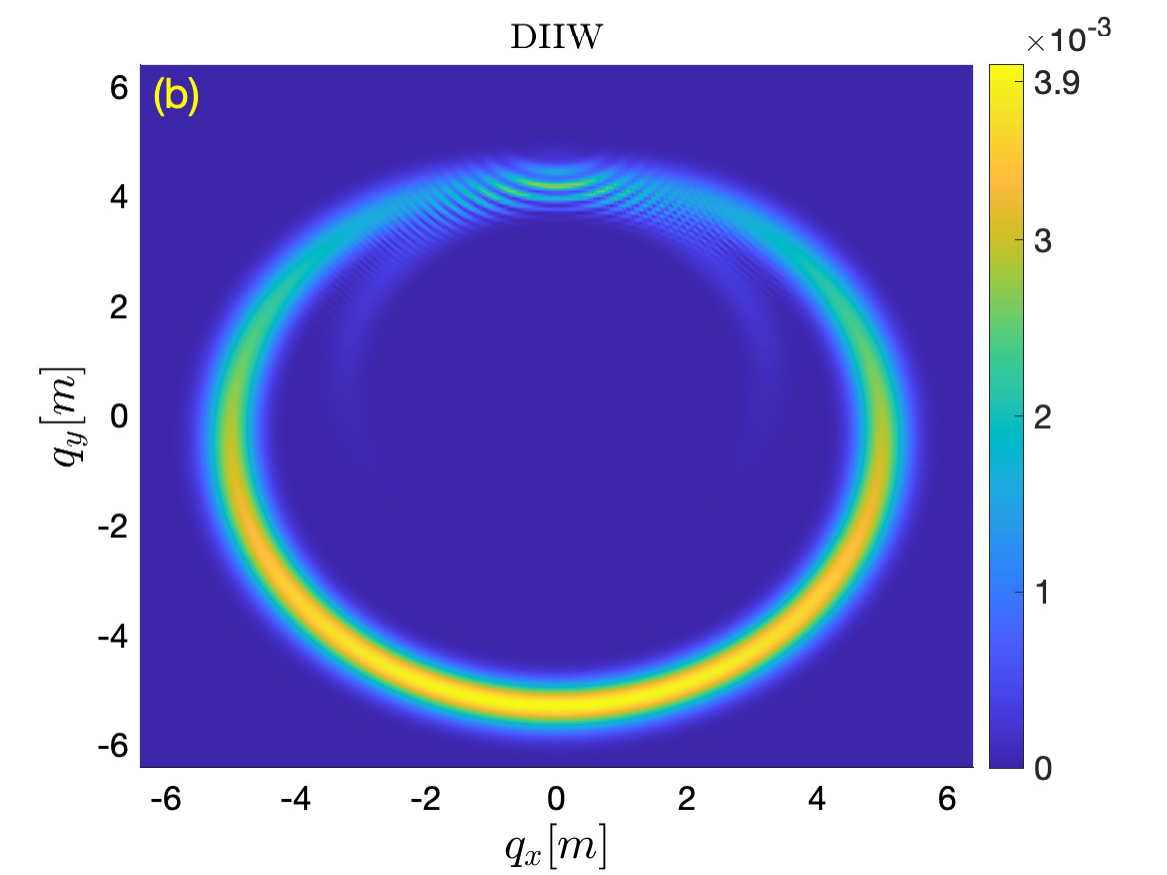}
\caption{Total electron momentum spectra under the CP electric field (where $q_z=0$). (a) and (b) are the numerical results of fermionic TLM and DHW methods. Other field parameters are $\varepsilon_{0}=0.5$, $N=5$ and $\omega_{0}=0.1 m$.
\label{fig:4}}
\end{figure}
The momentum spectra obtained from the fermionic TLM and DHW methods for the tunneling dominated process ($\gamma_{\omega}=0.2$)  are shown in Figs.~\ref{fig:4}(a) and (b).
We can mainly see one ring in the momentum spectra for both the fermionic TLM and DHW methods.
Moreover, one can also find that the radius of the ring in the momentum spectra in tunneling dominated process are even larger than that of the radius of the ring in the the momentum spectra in multiphoton dominated process for fermionic case. We further find that the two results from the fermionic TLM and DHW methods are almost the same, i.e., the ratio of the two momentum spectra $R_F$ is about $1.02$.
One can clearly find the interference effect in the momentum spectra.
The fermionic TLM is completely equivalent to the quantum mechanical scattering problem. Thus, the origin of the interference effect is the resonances in the scattering problem as the shape of the potential $-\omega^2_{\mathbf{q}}(t)$ changes with the momentum $\mathbf{q}$ \cite{Hebenstreit:2009km}. We want to stress that the approximate symmetry of the momentum spectra with respect to the $q_y=0$ disappears during the tunneling dominated process, but the exact symmetry for $q_x=0$ still exists.

\begin{figure}[ht!]\centering
\includegraphics[width=0.49\textwidth]{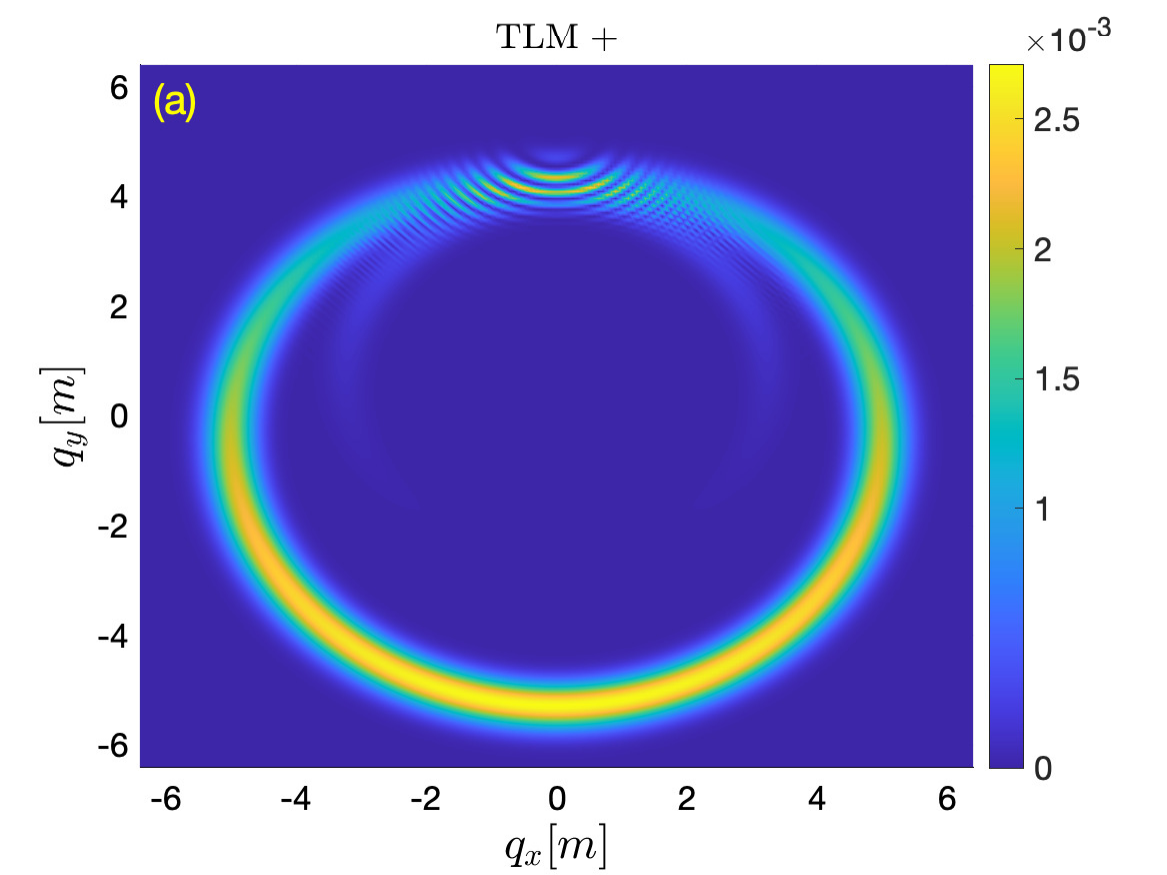}
\includegraphics[width=0.49\textwidth]{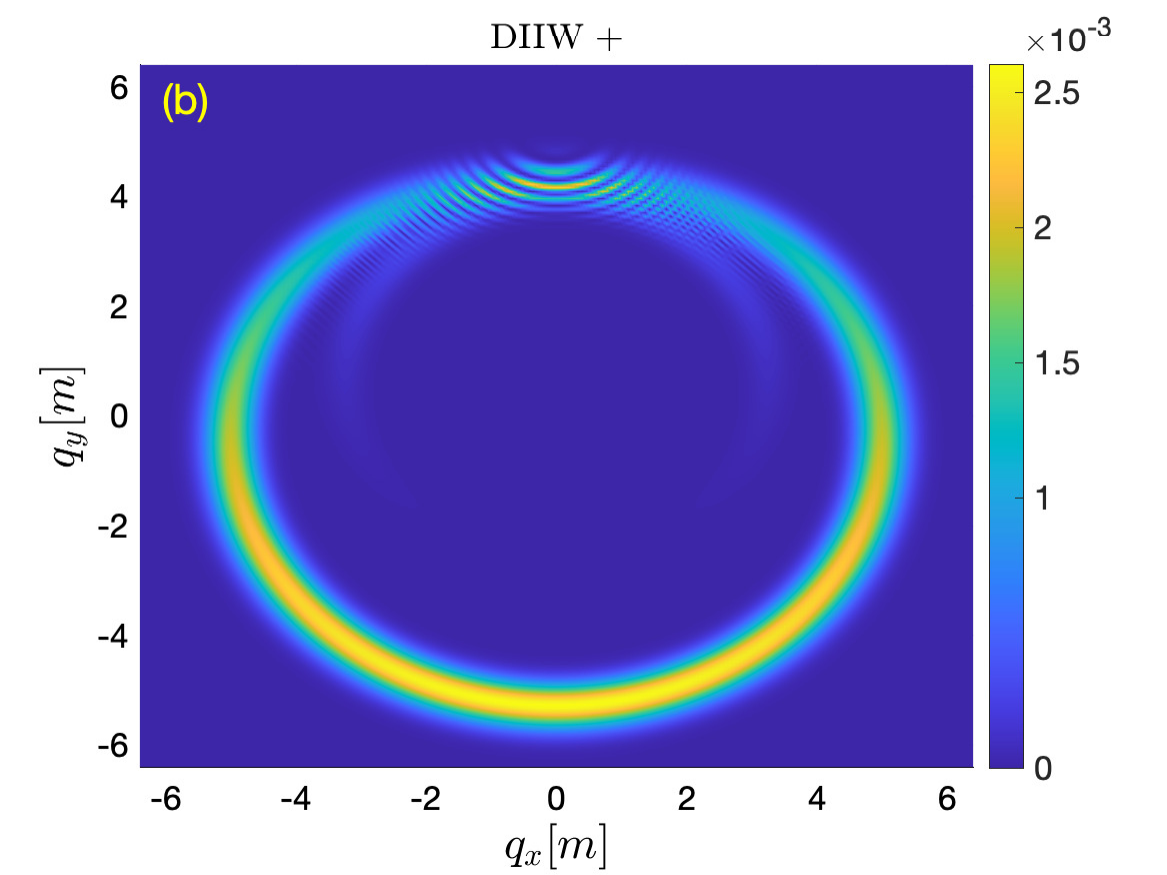}
\includegraphics[width=0.49\textwidth]{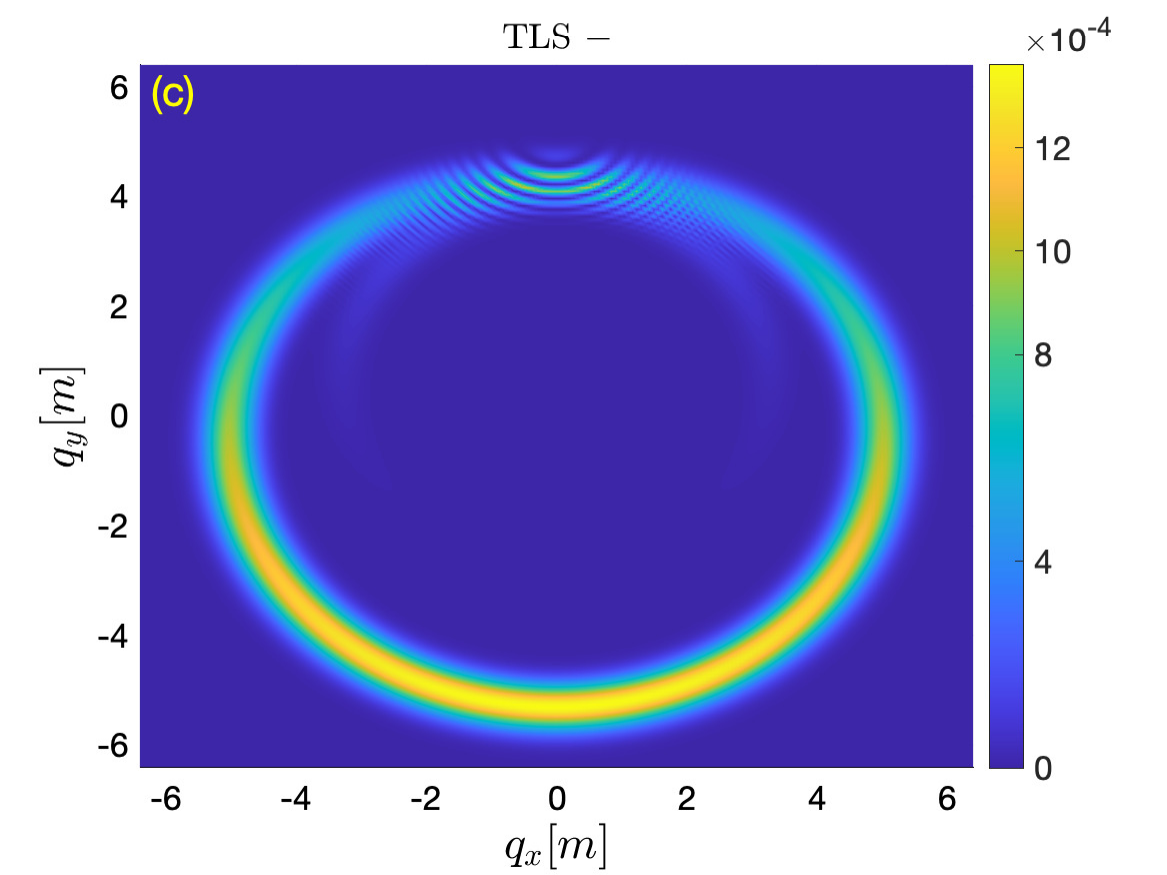}
\includegraphics[width=0.49\textwidth]{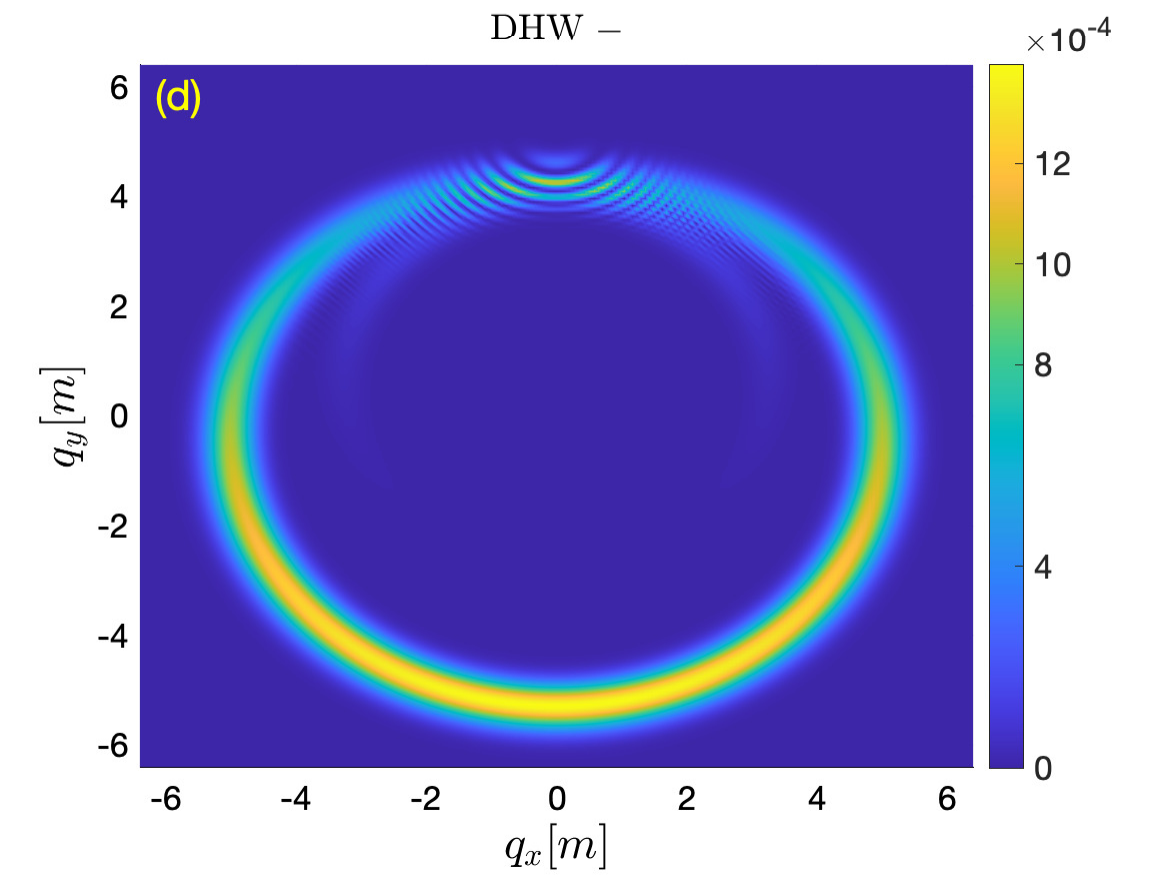}
\caption{Electron momentum spectra under the CP electric field with different method and spin states (where $q_z=0$). The first and second columns represent the fermionic TLM and DHW results respectively. The first and second rows represent the results of electron spin $+$ and $-$, respectively. Other field parameters are $\varepsilon_{0}=0.5$, $N=5$ and $\omega_{0}=0.1 m$.
\label{fig:5}}
\end{figure}

One can interpret why the electron momentum spectra of the two methods are almost the same.
It could be found that the spin-dependent momentum spectra for both the fermionic TLM and DHW methods are relatively similar in terms of the general trend, but there are slight differences yet, as shown in Figs.~\ref{fig:5}(a), (b), (c) and (d). For instance, one ring on the electron momentum spectra $f^{+}_{\rm TLM}$ and $f^{+}_{\rm DHW}$ can be obviously seen in Figs.~\ref{fig:5}(a) and (b). We also obtain $R^+_F=f^+_{\rm TLM}/f^+_{\rm DHW}\thickapprox 1.21$ and $R^-_F=f^-_{\rm TLM}/f^-_{\rm DHW}\thickapprox 0.96$. $f^{+}_{\rm TLM}$ is approximately two orders of magnitude larger than $f^{-}_{\rm TLM}$. As a result, the primary contribution to the electron momentum spectra in Figs.~\ref{fig:4}(a) and (b) originates from the spin $+$ electron momentum spectra in Figs.~\ref{fig:5}(a) and (b).

\begin{figure}[ht!]\centering
\includegraphics[width=0.49\textwidth]{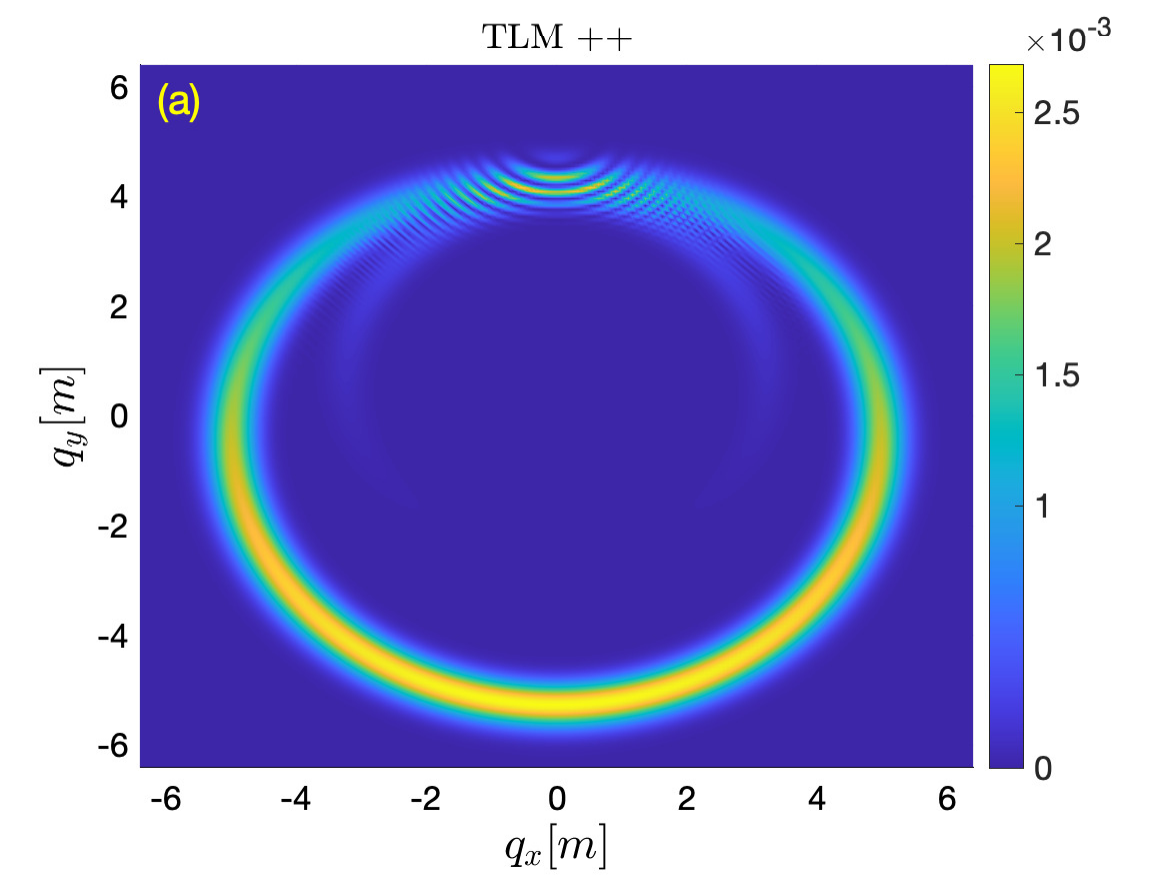}
\includegraphics[width=0.49\textwidth]{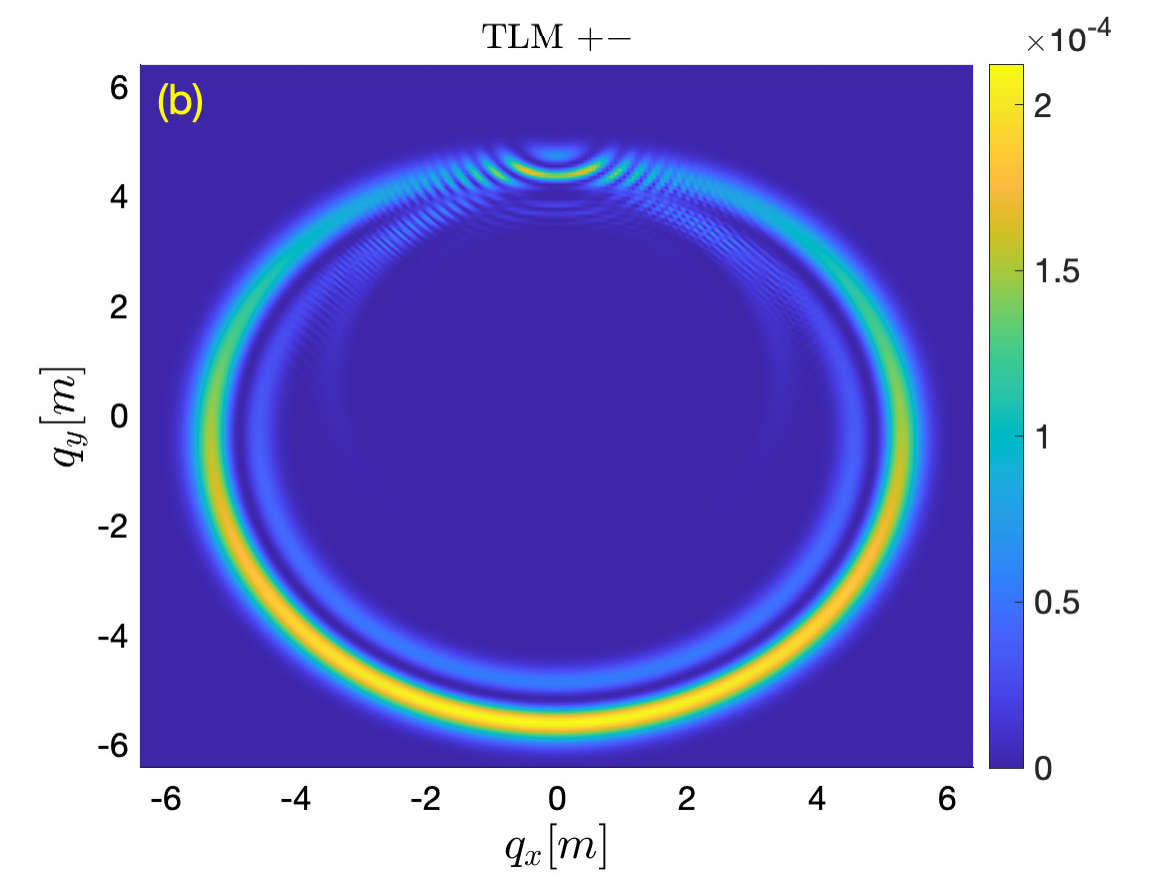}
\includegraphics[width=0.49\textwidth]{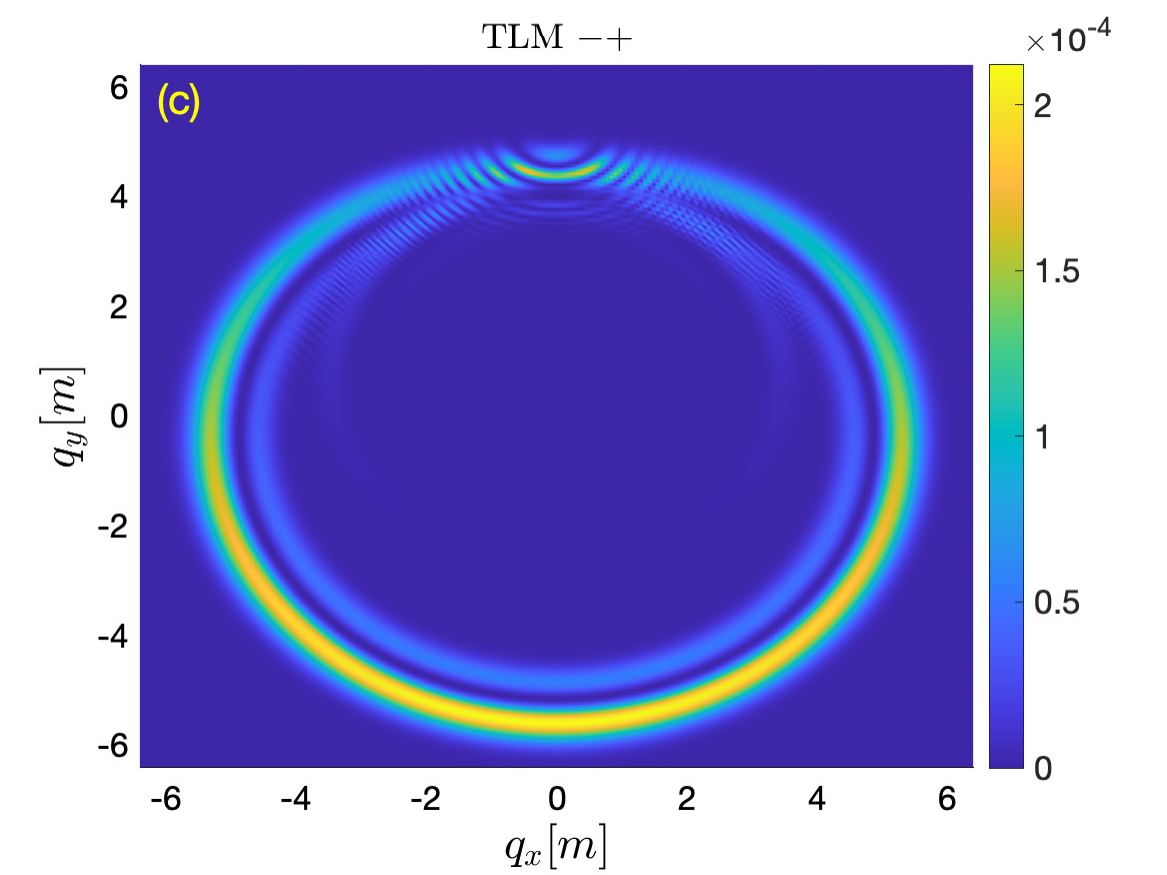}
\includegraphics[width=0.49\textwidth]{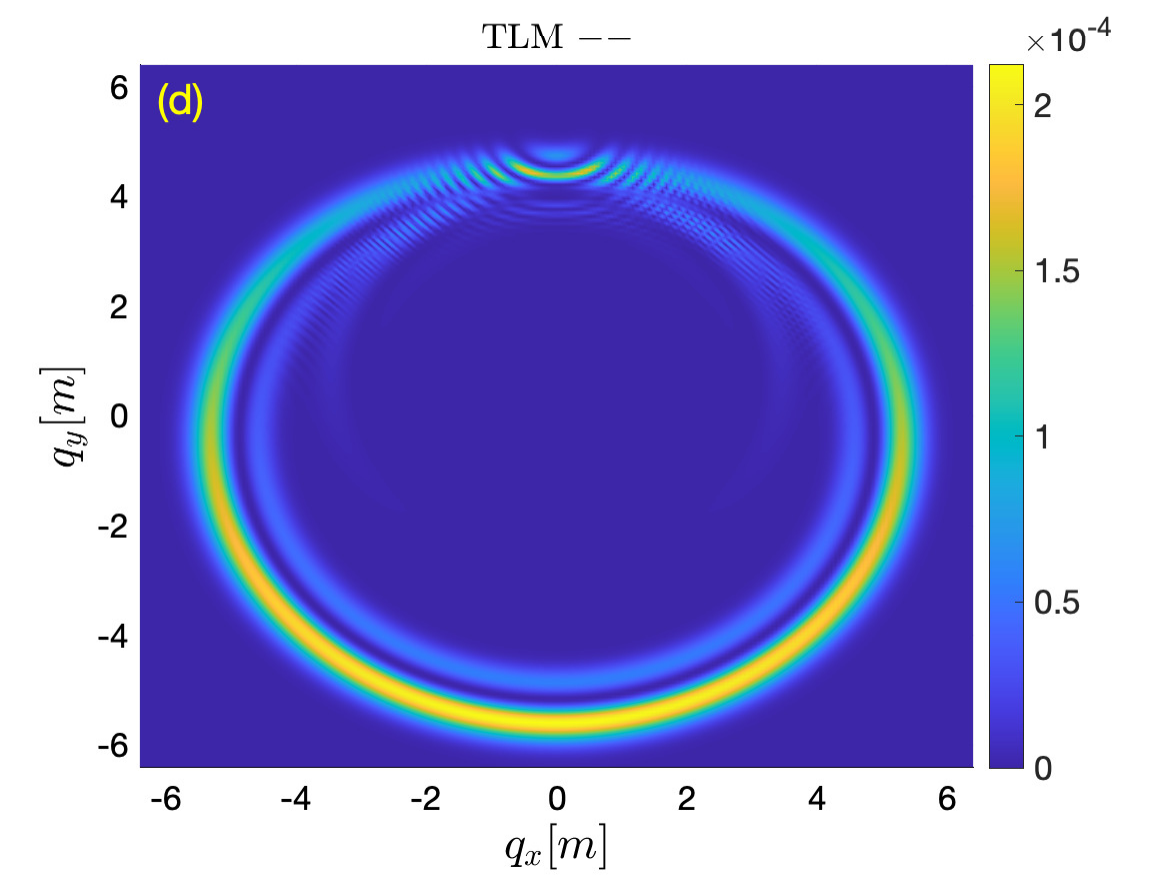}
\caption{Momentum spectra $f^{{\bf s}_{e^+}{\bf s}_{e^-}}_{\rm TLM}$ with different electron and positron spin states via fermionic TLM (where $q_z=0$). The first and second rows represent positron spin $+$ and $-$, respectively. The first and second columns represent the results of electron spin $+$ and $-$, respectively. Other field parameters are $\varepsilon_{0}=0.5$, $N=5$ and $\omega_{0}=0.1 m$.
\label{fig:6}}
\end{figure}

Let us interpret the spin-dependent momentum spectra. The full spin resolved momentum spectra $f^{{\bf s}_{e^+}{\bf s}_{e^-}}_{\rm TLM}$ are given in Figs.~\ref{fig:6}(a), (b), (c), and (d). We know $f^{+}_{\rm TLM}=f^{++}_{\rm TLM}+f^{-+}_{\rm TLM}$, this leads to the emergence of a single ring in the electron momentum spectrum, as shown in Fig.~\ref{fig:5}(a). Similarly, the electron spectrum $f^{-}_{\rm TLM}$, presented in Fig.~\ref{fig:5}(c), is the sum of all positron spin-dependent momentum spectra, namely $f^{+-}_{\rm TLM}+f^{--}_{\rm TLM}$. Consequently, a single ring is observable in the electron momentum spectrum, see Fig.~\ref{fig:5}(c).

In the end, we find that the momentum spectrum $f^{++}_{\rm TLM}$ is the largest for all alignments of electron and positron spin states in the tunneling dominated process. The physical interpretation is the same as in Sec.~\ref{sec:4A}.

\section{Summary}\label{sec:5}

In summary, we have formulated a generalized fully spin resolved fermionic TLM under arbitrarily time-dependent electric fields, i.e., one-dimensional fermionic TLM is generalized to three dimensional fermionic TLM. Our generalized TLM can provide momentum spectra dependent on both spin states of particle and anti-particle. We did not discuss scalar particle and anti-particle creation in the main text, but we discussed the validity of the scalar TLM by comparing with the FVHW method for different processes in the Appendix~\ref{ap:c}. We found that the results given by the scalar and fermionic TLM almost consistent with the FVHW and DHW formalisms, including the details of the momentum spectra, such as the interference effect of momentum spectra. Interestingly, it could be found that the momentum spectrum $f^{++}_{\rm TLM}$ is the greatest among all the spin alignment in the fermionic TLM.

Compared to the FVHW and DHW methods, TLM is simple, because it has only two differential equations to solve, while the FVHW and DHW methods have three or ten differential equations in a time-dependent external field. Additionally, we can apply the generalized TLM to investigate the vortex structures and the phase of the probability amplitude in the context of vacuum pair production.

\section{Acknowledgments}\label{sec:6}
We are grateful to LN Hu and QZ Lv for helpful discussions. This work was supported by the National Natural Science Foundation of China (NSFC) under Grant No. 12375240, No. 11935008 and No. 12233002. The computation was carried out at the HSCC of the Beijing Normal University. YFH was also supported by National SKA Program of China No. 2020SKA0120300, by the National Key R\&D Program of China (2021YFA0718500), and by the Xinjiang Tianchi Program.

\appendix

\section{Scalar two level model}\label{ap:a}
If we consider the scalar pair production in an arbitrary time-dependent arbitrarily electric field $\mathbf{E}(t)=(E_x(t), E_y(t), E_z(t))$, one can decompose the scalar field operator as
\begin{eqnarray}\label{A1}
&&\Phi(\mathbf{x},t)=\int d^3k\,  e^{i\mathbf{k}\cdot \mathbf{x}} \left(\phi_{\bf{k}}(t)a_{\bf{k}}+\phi^{*}_{\bf{k}}(t)b^{\dagger}_{-\bf{k}}\right),
\end{eqnarray}
where $a_{\bf{k}}$ and $b^{\dagger}_{-\bf{k}}$ satisfy standard bosonic commutation relations, i.e.,
\begin{eqnarray}\label{A2}
[a_{\bf{k}},a^\dagger_{\bf{k'}}]=[b_{\bf{k}},b^\dagger_{\bf{k'}}]=\delta^3(\bf{k}-\bf{k'}).
\end{eqnarray}
Our first aim is to solve the Klein-Gordon equation coupled to the external field
\begin{eqnarray}\label{A3}
[(i \partial-q A)^2-m^2)]\Phi(\mathbf{x},t)=0.
\end{eqnarray}
We can further rewrite the solution of the Klein-Gordon equation in the separable form
\begin{eqnarray}\label{A4}
\Phi(\mathbf{x},t)= \int \frac{d^3\mathbf{k}}{(2 \pi)^3} e^{i\mathbf{k}\cdot \mathbf{x}} \varPhi_{\mathbf{k}}(t).
\end{eqnarray}
After substituting Eq.~\eqref{A4} into Eq.~\eqref{A3}, we find that the function $\varPhi_{\mathbf{k}}(t)$ satisfies the harmonic oscillator equation
\begin{eqnarray}\label{A5}
\biggl[\frac{d^2}{dt^2} + \omega_{\mathbf{k}}^2(t)\biggr]\varPhi_{\mathbf{k}}(t)=0,
\end{eqnarray}
where $\omega_{\mathbf{k}}(t)=\sqrt{(\mathbf{k}-q\mathbf{A}(t))^2+m^2}$.

We now apply the Bogoliubov transformation\cite{Marinov:1977gq,Dumlu:2011rr} by defining $\alpha_{\bf{k}}(t)$ and $\beta_{\bf{k}}(t)$ as
\begin{align}\label{A6}
\phi_{\bf{k}}(t)&=\frac{\alpha_{\bf{k}}(t)}{\sqrt{2 \omega_{\mathbf{k}}(t)}}e^{-i\int^t d\tau \omega_{\mathbf{k}}(\tau)}+\frac{\beta_{\bf{k}}(t)}{\sqrt{2 \omega_{\mathbf{k}}(t)}}e^{i\int^t d\tau \omega_{\mathbf{k}}(\tau)},\\ \label{A7}
\dot{\phi}_{\bf{k}}(t)&=-i\omega_{\mathbf{k}}(t)\left(\frac{\alpha_{\bf{k}}(t)}{\sqrt{2 \omega_{\mathbf{k}}(t)}}e^{-i\int^t d\tau \omega_{\mathbf{k}}(\tau)}-\frac{\beta_{\bf{k}}(t)}{\sqrt{2 \omega_{\mathbf{k}}(t)}}e^{i\int^t d\tau \omega_{\mathbf{k}}(\tau)}\right).
\end{align}
By substituting Eq.~\eqref{A6} into Eq.~\eqref{A5} and comparing with Eq.~\eqref{A7}, we can easily obtain
\begin{eqnarray}\label{A8}
\dot{\alpha}_{\bf{k}}(t)&=&\frac{\dot{Q}_{\bf{k}}(t)}{2\omega_{\mathbf{k}}(t)}\beta_{\bf{k}}(t)e^{2i\int^t d\tau \omega_{\mathbf{k}}(\tau)},\\ \label{A9}
\dot{\beta}_{\bf{k}}(t)&=&\frac{\dot{Q}_{\bf{k}}(t)}{2\omega_{\mathbf{k}}(t)}\alpha_{\bf{k}}(t)e^{-2i\int^t d\tau \omega_{\mathbf{k}}(\tau)}.
\label{abdot}
\end{eqnarray}
The role of this Bogoliubov transformation is that it changes from time-independent bases of creation $a_{\bf k}$ and annihilation operators $b_{-{\bf k}}^\dagger$ to time-dependent bases of creation $\tilde{a}_{\bf k}(t)$ and annihilation operators $\tilde{b}_{-{\bf k}}^\dagger(t)$ by performing a linear transformation \cite{Kluger:1998bm,Dumlu:2011rr} as
\begin{eqnarray}\label{A10}
\begin{pmatrix}
\tilde{a}_{\bf k}(t)\cr
\tilde{b}_{-{\bf k}}^\dagger(t)
\end{pmatrix}
=\begin{pmatrix}
\alpha_{\bf k} & \beta_{\bf k}^*\cr
\beta_{\bf k} & \alpha_{\bf k}^*
\end{pmatrix}
\begin{pmatrix}
a_{\bf k}\cr
b^\dagger_{-{\bf k}}
\end{pmatrix},
\end{eqnarray}
where the bosonic commutation relations are preserved by $|\alpha_{\bf{k}}(t)|^2 -|\beta_{\bf{k}}(t)|^2=1$.

Now we introduce the new useful coefficients \cite{Akkermans:2011yn}
\begin{eqnarray}\label{A11}
c^{(1)}_{\bf{k}}(t)&=&\beta_{\bf{k}}(t)e^{-i\int^t d\tau \omega_{\mathbf{k}}(\tau)},\\ \label{A12}
c^{(2)}_{\bf{k}}(t)&=&\alpha_{\bf{k}}(t)e^{i\int^t d\tau \omega_{\mathbf{k}}(\tau)},
\end{eqnarray}
and after taking the derivatives of time on the above equations, we can obtain the scalar TLM
\begin{equation}\label{A13}
i\frac{d}{dt}\begin{bmatrix}
c_{\bf{k}}^{(1)}(t)\\
c_{\bf{k}}^{(2)}(t)
\end{bmatrix}
 =
 \begin{pmatrix} \omega_{\bf{k}}(t) & -i\Omega_{\bf{k}}(t) \cr -i\Omega_{\bf{k}}(t) & -\omega_{\bf{k}}(t) \end{pmatrix}
 \begin{bmatrix}
c_{\bf{k}}^{(1)}(t)\\
c_{\bf{k}}^{(2)}(t)
\end{bmatrix},
\end{equation}
 where
\begin{equation}\label{A14}
\Omega_{\bf{k}}(t)=-\frac{\dot{\omega}_{\bf{k}}(t)}{2\omega_{\bf{k}}(t)}=-\frac{q\mathbf{E}(t)\cdot(\mathbf{k}-q\mathbf{A}(t))}{2\omega_{\bf{k}}^2(t)}=-\frac{q\mathbf{E}(t)\cdot{\bf p}(t)}{2\omega_{\bf{k}}^2(t)},
\end{equation}
and the initial conditions are $c_{\bf{k}}^{(1)}(t_0)=1$ and $c_{\bf{k}}^{(2)}(t_0)=0$.

The electron momentum distribution can be calculated by using the coefficient $c_{\bf{k}}^{(2)}$ at $t = +\infty$ as
 \begin{eqnarray}\label{A15}
f_{\bf k}=2|c_{\bf{k}}^{(2)}(t=+\infty)|^2.
\end{eqnarray}

\section{Feshbach-Villars-Heisenberg-Wigner formalism}\label{ap:b}
In this appendix, we will briefly review the equal-time Feshbach-Villars-Heisenberg-Wigner (FVHW) formalism which corresponds to the Wigner function for scalar QED. The Klein-Gordon equation in an electromagnetic field is our starting point
\begin{align}\label{D1}
(D^\mu D_\mu+m^2)\Phi (\mathbf{x},t)=0,
\end{align}
where $D_\mu=\partial_\mu+ieA_\mu(\mathbf{x},t)$ is the covariant derivative, $e$, $m$ and $A_\mu(\mathbf{x},t)$ are the electron charge, mass and four-dimensional vector potential of the external field respectively. Equation (\ref{D1}) can be written in the Schr\"{o}dinger form
\begin{align}\label{D2}
i\frac{\partial}{\partial t}\Phi (\mathbf{x},t)=\hat{\mathbbm{H}}(\mathbf{x},t)\Phi (\mathbf{x},t).
\end{align}
We use the Feshbach-Villars representation \cite{Feshbach:1958wv}
\begin{eqnarray}\label{D3}
\Phi &=&\left(
              \begin{array}{c}
              \xi \\
              \eta \\
              \end{array}
              \right),\\ \label{D4}
\xi&=&\frac{1}{2}\Big(\psi+\frac{i}{m}\frac{\partial\psi}{\partial t}-\frac{qA^0}{m}\psi\Big),\\ \label{D5}
\eta&=&\frac{1}{2}\Big(\psi-\frac{i}{m}\frac{\partial\psi}{\partial t}+\frac{qA^0}{m}\psi\Big),
\end{eqnarray}
and the two-component nonhermitian Hamiltonian
\begin{align}\label{D6}
\hat{\mathbbm{H}}&=\frac{(\hat{p}-e\mathbf{A})^2}{2m}\mathbbm{a}+m\mathbbm{b}+qA^0\mathbbm{1},\\ \label{D7}
\mathbbm{a}&=\sigma_3+i\sigma_2=
\left(
\begin{array}{cc}
    1 & 1 \\
    -1 & -1 \\
\end{array}
\right),\\  \label{D8}
\mathbbm{b}&=\sigma_3=\left(
  \begin{array}{cc}
    1 & 0 \\
    0 & -1 \\
  \end{array}
\right),\\  \label{D9}
\mathbbm{1}&=\left(
  \begin{array}{cc}
    1 & 0 \\
    0 & 1 \\
  \end{array}
\right),
\end{align}
where $\hat{p}=-i\nabla$.

The equal-time density operator for scalar QED, which is an equal-time anti-commutator constructed by two Feshbach-Villars field operators in the Heisenberg picture, can be expressed as \cite{Best:1993wq}
\begin{align}\label{D10}
\hat{\mathcal{C}}_{scalar}(\mathbf{x}_1,\mathbf{x}_2,t)\equiv \exp\Big(-ie\int_{\mathbf{x}_2}^{\mathbf{x}_1} d\mathbf{x}'\cdot\mathbf{A}(\mathbf{x}',t)\Big) \Big\{\Phi^\dagger (\mathbf{x}_1,t),{\Phi }(\mathbf{x}_2,t)\Big\},
\end{align}
where the exponential factor on the right side of this equation is the Wilson line factor, which is introduced for maintaining gauge invariance. For the center of mass coordinate $\mathbf{x}=(\mathbf{x}_1+\mathbf{x}_2)/2$ and the relative coordinate $\mathbf{r}=\mathbf{x}_1-\mathbf{x}_2$, the density operator becomes
\begin{align}\label{D11}
\hat{\mathcal{C}}_{scalar}(\mathbf{x},\mathbf{r},t)=\exp\Big(-ie\int_{-1/2}^{1/2}d\lambda\,\mathbf{r}\cdot\mathbf{A}(\mathbf{x}+\lambda\mathbf{r},t)\Big) \Big\{\Phi^\dagger \Big(\mathbf{x}+\frac{\mathbf{r}}{2},t\Big),\Phi \Big(\mathbf{x}-\frac{\mathbf{r}}{2},t\Big)\Big\}.
\end{align}

After performing the Fourier transform on the vacuum expectation value of the density operator $\hat{\mathcal{C}}_{scalar}(\mathbf{x},\mathbf{r},t)$ with respect to the relative coordinate $\mathbf{r}$, we can obtain the Wigner function \cite{Zhuang:1995pd}
\begin{align}\label{D12}
\mathcal{W}_{scalar}(\mathbf{x},\mathbf{p},t)\equiv \frac{1}{2}\int{d^3 \mathbf{r}\,
\langle\mathrm{vac}|\hat{\mathcal{C}}_{scalar}(\mathbf{x},\mathbf{r},t)|\mathrm{vac}\rangle e^{-i\mathbf{p}\cdot\mathbf{r}}},
\end{align}
where $|\mathrm{vac}\rangle$ is the vacuum state in the Heisenberg picture and $\mathbf{p}$ is the kinetic momentum. Taking the time derivative of Eq.~(\ref{D12}) and employing Eqs.~\ref{D3}-\ref{D5}, the equation of motion for the Wigner function can be obtained as \cite{Zhuang:1995pd}
\begin{align}\label{D13}
\begin{split}
iD_t\mathcal{W}_{scalar}=&-i\frac{\mathbf{P}\cdot\mathbf{D}}{2m}(\mathbbm{a}\mathcal{W}_{scalar}
+\mathcal{W}_{scalar}\mathbbm{a}^\dagger)+\frac{4\mathbf{P}^2-\mathbf{D}^2}{8m}(\mathbbm{a}\mathcal{W}_{scalar}-\mathcal{W}_{scalar}\mathbbm{a}^\dagger)\\
&+m(\mathbbm{b}\mathcal{W}_{scalar}-\mathcal{W}_{scalar}\mathbbm{b}),
\end{split}
\end{align}
where $D_t$, $\mathbf{D}$ and $\mathbf{P}$ denote the non-local pseudo-differential operators:
\begin{eqnarray}\label{D14}
D_t&=&\frac{\partial}{\partial t}+e\int_{-1/2}^{1/2}{d\lambda\,\mathbf{E}\Big(\mathbf{x}
+i\lambda\frac{\partial}{\partial \mathbf{p}},t\Big)\cdot \frac{\partial}{\partial \mathbf{p}}}, \nonumber \\
\mathbf{D}&=&\nabla+e\int_{-1/2}^{1/2}{d\lambda\,\mathbf{B}\Big(\mathbf{x}
+i\lambda\frac{\partial}{\partial \mathbf{p}},t\Big)\times\frac{\partial}{\partial \mathbf{p}}}, \\
\mathbf{P}&=&\mathbf{p}-ie\int_{-1/2}^{1/2}{d\lambda\,\lambda\,\mathbf{B}\Big(\mathbf{x}
+i\lambda\frac{\partial}{\partial \mathbf{p}},t\Big)\times\frac{\partial}{\partial \mathbf{p}}}. \nonumber
\end{eqnarray}
Note that the Hartree approximation is used here, therefore, the electromagnetic field is treated as a C-number field instead of a Q-number one.

In order to obtain the transport equations in scalar QED, we can expand the matrix-valued Wigner functions in terms of the Feshbach-Villars spinors $\{\mathbbm{1},\sigma_{i=\{1,2,3\}}\}$ and $4$ real FVHW functions $\chi^{\mu=\{0,1,2,3\}}(\mathbf{x},\mathbf{p},t)$ \cite{Zhuang:1995pd}
\begin{equation}\label{D15}
  \mathcal{W}_{scalar}(\mathbf{x},\mathbf{p},t)=\frac{1}{2}\left(\chi^0\mathbbm{1}+\chi^1\sigma_1
  +\chi^2\sigma_2+\chi^3\sigma_3\right).
\end{equation}
We obtain a set of coupled partial integro-differential equations for the equal-time FVHW functions after inserting Eq.~(\ref{D15}) into Eq.~(\ref{D13}) and comparing the coefficients of the Feshbach-Villars spinors \cite{Zhuang:1995pd}
\begin{eqnarray}\label{D16}
D_t\chi^0\!&=&\!\frac{4\mathbf{P}^2-\mathbf{D}^2}{4m}\chi^2
-\frac{\mathbf{P}\cdot\mathbf{D}}{m}\chi^3,\\ \label{D17}
D_t\chi^1\!&=&\!-\Big(\frac{4\mathbf{P}^2-\mathbf{D}^2}{4m}+2m\Big)\chi^2
+\frac{\mathbf{P}\cdot\mathbf{D}}{m}\chi^3,\\ \label{D18}
D_t\chi^2\!&=&\!\frac{4\mathbf{P}^2-\mathbf{D}^2}{4m}\chi^0+
\Big(\frac{4\mathbf{P}^2-\mathbf{D}^2}{4m}+2m\Big)\chi^1, \;\\ \label{D19}
D_t\chi^3\!&=&\!-\frac{\mathbf{P}\cdot\mathbf{D}}{m}(\chi^0+\chi^1).
\end{eqnarray}
The above equations are called the FVHW formalism. Some of the FVHW functions can be given an obvious physical interpretation by the expression of conservation laws for some physically observable quantities. For instance, $\mathcal{Q}$, $\mathcal{J}$, $\mathcal{E}$ and $\mathcal{P}$ are the total charge, current, energy and linear momentum respectively, i.e.,
\begin{equation}\label{D20}
  \frac{d}{dt}\left\{\mathcal{Q};\mathcal{J};\mathcal{E};\mathcal{P}\right\}=0 \ ,
\end{equation}
with
\begin{eqnarray}\label{D21}
\mathcal{Q} \!&=&\! e\int d\Gamma\, \chi^3(\mathbf{x},\mathbf{p},t), \\ \label{D22}
\mathcal{J} \!&=&\! q\int d\Gamma\, \frac{\mathbf{p}}{m}[\chi^0(\mathbf{x},\mathbf{p},t)+\chi^1(\mathbf{x},\mathbf{p},t)], \\ \label{D23}
\mathcal{E} \!&=&\! \int d\Gamma\, \Big[\Big(\frac{\mathbf{p}^2}{2m}+m\Big)\chi^0(\mathbf{x},\mathbf{p},t)
+\frac{\mathbf{p}^2}{2m}\chi^1(\mathbf{x},\mathbf{p},t)\Big] \nonumber\\ \label{D24}
&&+\frac{1}{2}\int d^3x\, [\mathbf{E}^2(\mathbf{x},t)+\mathbf{B}^2(\mathbf{x},t)], \\ \label{D25}
\mathcal{P} \!&=&\! \int d\Gamma\, \mathbf{p}\,\chi^3(\mathbf{x},\mathbf{p},t)+\int d^3x\, \mathbf{E}(\mathbf{x},t)\times\mathbf{B}(\mathbf{x},t),
\end{eqnarray}
where $d\Gamma=d^3x\,d^3p/(2\pi)^3$ is the phase-space volume. According to the above expressions, $m \chi^0(\mathbf{x},\mathbf{p},t)$ may be associated with a mass density, $\frac{\mathbf{p}}{m}[\chi^0(\mathbf{x},\mathbf{p},t)+\chi^1(\mathbf{x},\mathbf{p},t)]$ with a current density and $q \chi^3(\mathbf{x},\mathbf{p},t)$ with a charge density. However, the quantity $\chi^2(\mathbf{x},\mathbf{p},t)$ has no obvious physical interpretation yet.

For an arbitrary spatially homogeneous time-dependent field (the temporal gauge $A_0=0$ can be chosen), the non-local operators Eq.~(\ref{D14}) become local ones:
\begin{eqnarray} \label{D26}
  D_t&=& \frac{\partial}{\partial  t} + e \mathbf{E}(t)\cdot\frac{\partial}{\partial \mathbf{p}} , \nonumber \\
  \mathbf{D}&=& \nabla, \\
  \mathbf{P}&=&\mathbf{p}. \nonumber
\end{eqnarray}
The vacuum initial conditions can be obtained as
\begin{eqnarray}\label{initialvalues}\label{D27}
\chi_{\mathrm{\rm vac}}^0(\mathbf{p})&=&\frac{1}{2}\Big[\frac{m}{\omega(\mathbf{p})}
+\frac{\omega(\mathbf{p})}{m}\Big]\ , \nonumber \\
\chi_{\mathrm{\rm vac}}^1(\mathbf{p})&=&\frac{1}{2}\Big[\frac{m}{\omega(\mathbf{p})}
-\frac{\omega(\mathbf{p})}{m}\Big]\ , \\
\chi_{\mathrm{\rm vac}}^2(\mathbf{p})&=&\chi_{\mathrm{\rm vac}}^3(\mathbf{p})=0,\nonumber
\end{eqnarray}
where the subscript `$\rm vac$' represents the vacuum initial condition, $\omega(\mathbf{p})=\sqrt{\mathbf{p}^2+m^2}$ is the electron energy.  Therefore, according to the vacuum initial conditions Eq. (\ref{initialvalues}), we find that the function $\chi^3(\mathbf{p},t)\equiv0$. Finally, we obtain the FVHW formalism as
\begin{equation}\label{D28}
\left[\frac{\partial}{\partial t}+q\mathbf{E}(t)\cdot\frac{\partial}{\partial \mathbf{p}}\right]\left\{
                    \begin{array}{c}
                      \chi^0 \\
                      \chi^1 \\
                      \chi^2 \\
                    \end{array}
                  \right\}(\mathbf{p},t)
=\mathcal{M}(\mathbf{p})\left\{
                    \begin{array}{c}
                      \chi^0 \\
                      \chi^1 \\
                      \chi^2 \\
                    \end{array}
                  \right\}(\mathbf{p},t),
\end{equation}
where
\begin{equation}\label{D29}
 \mathcal{M}(\mathbf{p})=\left(\begin{array}{ccc}
 0&0&\frac{\mathbf{p}^2}{m}\\
 0&0&-\frac{\mathbf{p}^2}{m}-2m\\
 \frac{\mathbf{p}^2}{m}&\frac{\mathbf{p}^2}{m}+2m&0
\end{array}\right).
\end{equation}

The partial differential equations Eq.~(\ref{D28}) can be simplified as an ordinary differential equation system after replacing $\mathbf{p}(t)$ by $\mathbf{k}-q\mathbf{A}(t)$ as:
\begin{eqnarray}\label{D29}
\frac{d}{d t} \chi^0(\mathbf{k},t)\!&=&\!\frac{\mathbf{p}^2(t)}{m}\chi^2(\mathbf{k},t),\nonumber\\
\frac{d}{d t} \chi^1(\mathbf{k},t)\!&=&\!-\Big[\frac{\mathbf{p}^2(t)}{m}+2m\Big]\chi^2(\mathbf{k},t),\\
\frac{d}{d t} \chi^2(\mathbf{k},t)\!&=&\!\frac{\mathbf{p}^2(t)}{m}\chi^0(\mathbf{k},t)+
\Big[\frac{\mathbf{p}^2(t)}{m}+2m\Big]\chi^1(\mathbf{k},t). \nonumber \,\,\,\,\quad
\end{eqnarray}

If we define three auxiliary quantities as
\begin{eqnarray}\label{D30}
\widetilde{\chi}^1\!&=&\!\Big[\frac{\mathbf{p}^2(t)}{2m\omega(\mathbf{k},t)}
+\frac{m}{\omega(\mathbf{k},t)}\Big]\chi^0
+\frac{\mathbf{p}^2(t)}{2m\omega(\mathbf{k},t)}\chi^1, \nonumber\quad\\
\mathcal{G}\!&=&\!\frac{\mathbf{p}^2(t)}{2m\omega(\mathbf{k},t)}\chi^0
+\Big[\frac{\mathbf{p}^2(t)}{2m\omega(\mathbf{k},t)}
+\frac{m}{\omega(\mathbf{k},t)}\Big]\chi^1, \quad\\
\mathcal{H}\!&=&\!\chi^2,\nonumber
\end{eqnarray}
then the ODE system Eq.~(\ref{D29}) can be transformed into the quantum Vlasov equation (QVE) \cite{Kluger:1998bm,Schmidt:1998vi}
\begin{eqnarray}\label{D31}
  \frac{d}{dt}\mathcal{F}(\mathbf{k},t)\!\!&=&\!\!\frac{1}{2}W(\mathbf{k},t)\,\mathcal{G}(\mathbf{k},t) \ , \nonumber \\
  \!\frac{d}{dt}\mathcal{G}(\mathbf{k},t)\!\!&=&\!\!W(\mathbf{k},t)[1+2\mathcal{F}(\mathbf{k},t)]
  -2\omega(\mathbf{k},t)\,\mathcal{H}(\mathbf{k},t), \;\;\quad
  \\
  \frac{d}{dt}\mathcal{H}(\mathbf{k},t)\!\!&=&\!\!2\omega(\mathbf{k},t)\,\mathcal{G}(\mathbf{k},t) \ , \nonumber
\end{eqnarray}
where
\begin{align}\label{D32}
\mathcal{F}(\mathbf{k},t)&=\frac{\widetilde{\chi}^1(\mathbf{k},t)-1}{2},\\ \label{D33}
W(\mathbf{k},t)&=\frac{q\mathbf{E}(t)\cdot\mathbf{p}(t)}{\omega^2(\mathbf{k},t)}.
\end{align}
The initial conditions are
\begin{equation}\label{D34}
  \mathcal{F}(\mathbf{k},t\rightarrow-\infty)=\mathcal{G} (\mathbf{k},t\rightarrow-\infty)=\mathcal{H}(\mathbf{k},t\rightarrow-\infty)=0.
\end{equation}
The number density of created real boson pairs can be calculated by integrating the momentum distribution function $\mathcal{F}(\mathbf{k},t\rightarrow\infty)$ with respect to the momentum $\mathbf{k}$:
\begin{equation}\label{numberdensity}
 n(t\rightarrow\infty)=\int \frac{d^3k}{(2\pi)^3}\mathcal{F}(\mathbf{k},t\rightarrow\infty).
\end{equation}
Here, we want to stress that the FVHW formalism Eq.~\eqref{D27} for a time-dependent electric field is fully equivalent to the multi-dimensional scalar QVE Eq.~\eqref{D31} for arbitrarily electric fields.

\section{scalar pair production}\label{ap:c}
Here, we investigate the validity of scalar TLM by comparing the FVHW method.
In order to comprehensively verify the consistency of these two methods, we will discuss the cases of multiphoton ($\gamma_{\omega} \gg 1$) and tunneling ($\gamma_{\omega} \ll 1$) processes separately, where the Keldysh parameters are defined as $\gamma_{\omega} = \frac{m\omega_{0}}{e\varepsilon_{0}E_{cr}}$ \cite{Blinne:2016yzv,Kohlfurst:2019mag}.

\subsection{Multi-photon dominated process}\label{ap:cA}

The virtual electron in vacuum can absorb photons from the background field and obtain energy to produce real electrons.
Therefore, the energy of an electron can be determined by the number of photons absorbed and the photon energy.
For scalar pair creation, we will use a background field with field strength of $\varepsilon_{0} = 0.1$, field frequency of $\omega_0 =  0.99 m$ and the number of cycles $N$ = 5, respectively.
The Keldysh parameter is $\gamma_{\omega} = 9.9$, indicating that the multiphoton process dominates.
\begin{figure}[ht!]\centering
\includegraphics[width=0.49\textwidth]{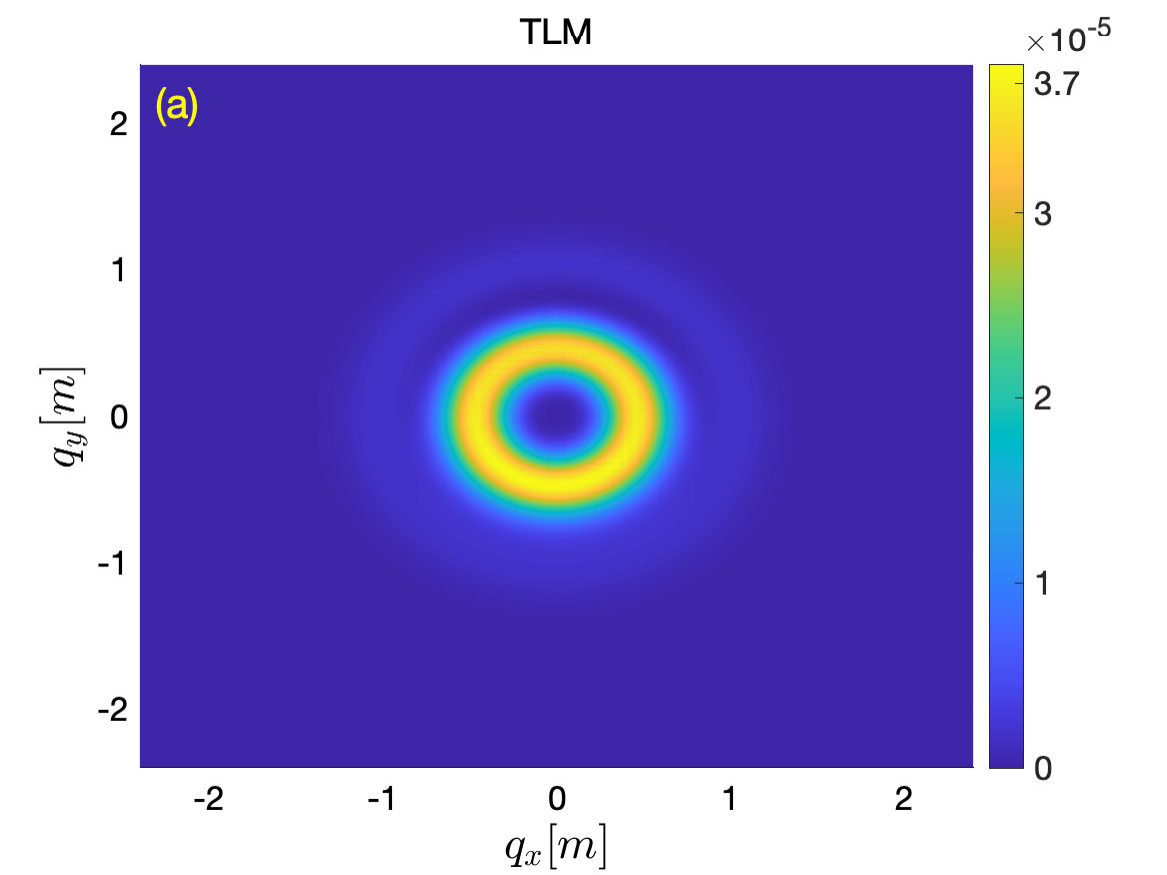}
\includegraphics[width=0.49\textwidth]{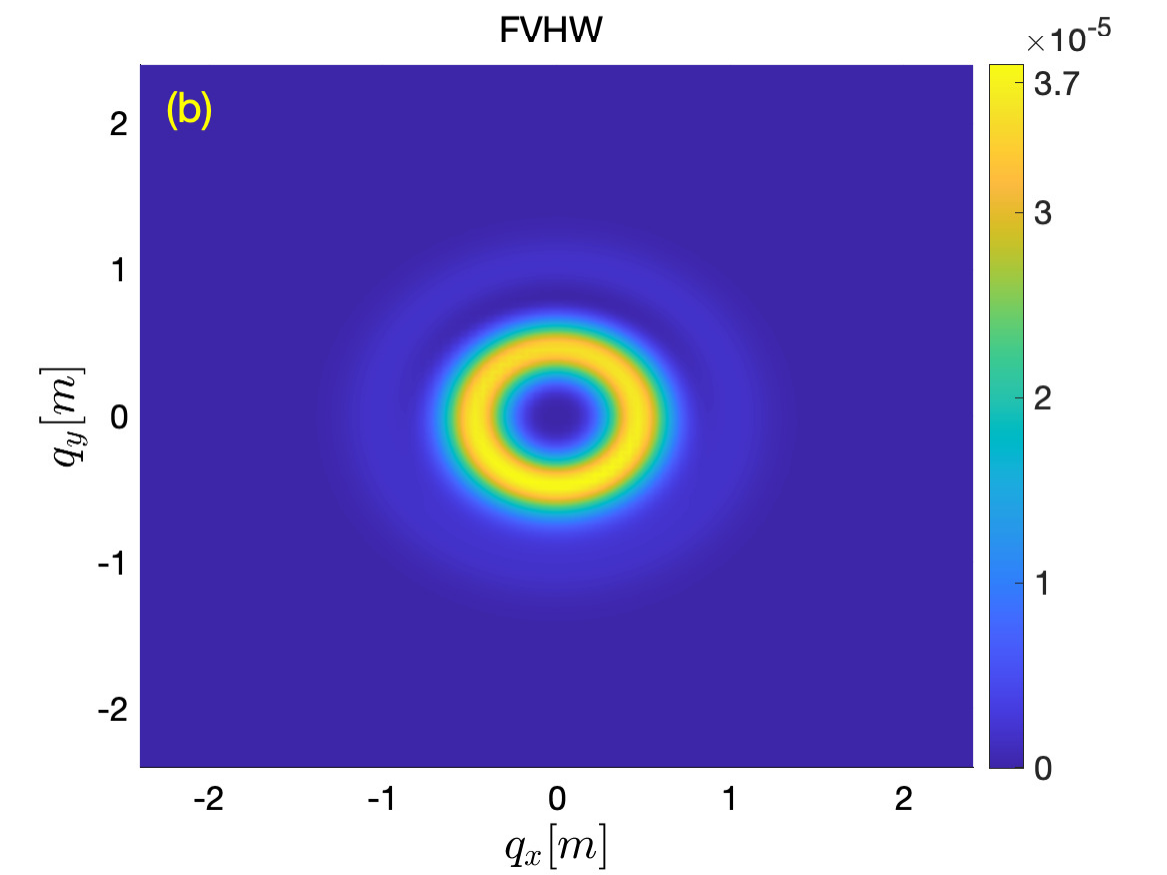}
\caption{Electron momentum spectra under the CP electric field (where $q_z=0$). (a) and (b) are the numerical results of scalar TLM and FVHW methods. Other field parameters are $\varepsilon_{0}=0.1$, $N=5$ and $\omega_{0}=0.99 m$.
\label{fig:10}}
\end{figure}
The momentum spectra of the scalar TLM and FVHW methods are shown in Figs.~\ref{fig:10}(a) and (b).
We can mainly find two concentric rings, which indicate that is typical of multiphoton processes. If we assume that it is a large pulse, the effective mass may be written as $m_{*}\thickapprox m\sqrt{1+\varepsilon_0^2 m^2/4\omega_0^2}$.
Consequently, the photon number could be estimated by $n=2\sqrt{m_{*}^2+q_{n}^2}/\omega_{0}$, i.e., $q_2\thickapprox 0.47m$ and $q_3\thickapprox 1.12m$ for $n=2$ and $n=3$ photons. Therefore, the ring structure comes from multiphoton dominated process by absorbing $n=2$ and $n=3$ photons.

Here we want to stress that the momentum spectrum is mainly related to the total energy of electron $\omega_{\mathbf{q}}(t)=\sqrt{m^2 + [q_x - e A_{x}(t)]^2+ [q_y - e A_{y}(t)]^2}$, where the vector potentials $A_{x}(t)$ and $A_{y}(t)$ are odd and even functions with respect to $t$, respectively. Therefore, $\omega_{\mathbf{q}}(t)$ stays invariant when replacing $t \rightarrow -t$ and $q_x \rightarrow -q_x$. This indicates that all momentum spectra in our work are exactly symmetric with respect to $q_{x}=0$. For the multiphoton dominated process, $q^{max}_{2,y}\gg A^{max}_{y}(t)$ so that $\omega_{\mathbf{q}}(t)\thickapprox \sqrt{m^2 + [q_x - e A_{x}(t)]^2+ q_y^2}$ stays invariant. Hence, the momentum spectra for multiphoton dominated process are approximately symmetric with respect to $q_{y}=0$.

The two results from the scalar TLM and FVHW methods are substantially the same, i.e.,
the ratio of the two momentum spectra $R_B=f_{\rm TLM}/f_{FVHW}$ is about $1$. It shows that the two methods are consistent in the multiphoton dominated process.

\subsection{Tunneling dominated process}\label{ap:cB}
\begin{figure}[ht!]\centering
\includegraphics[width=0.49\textwidth]{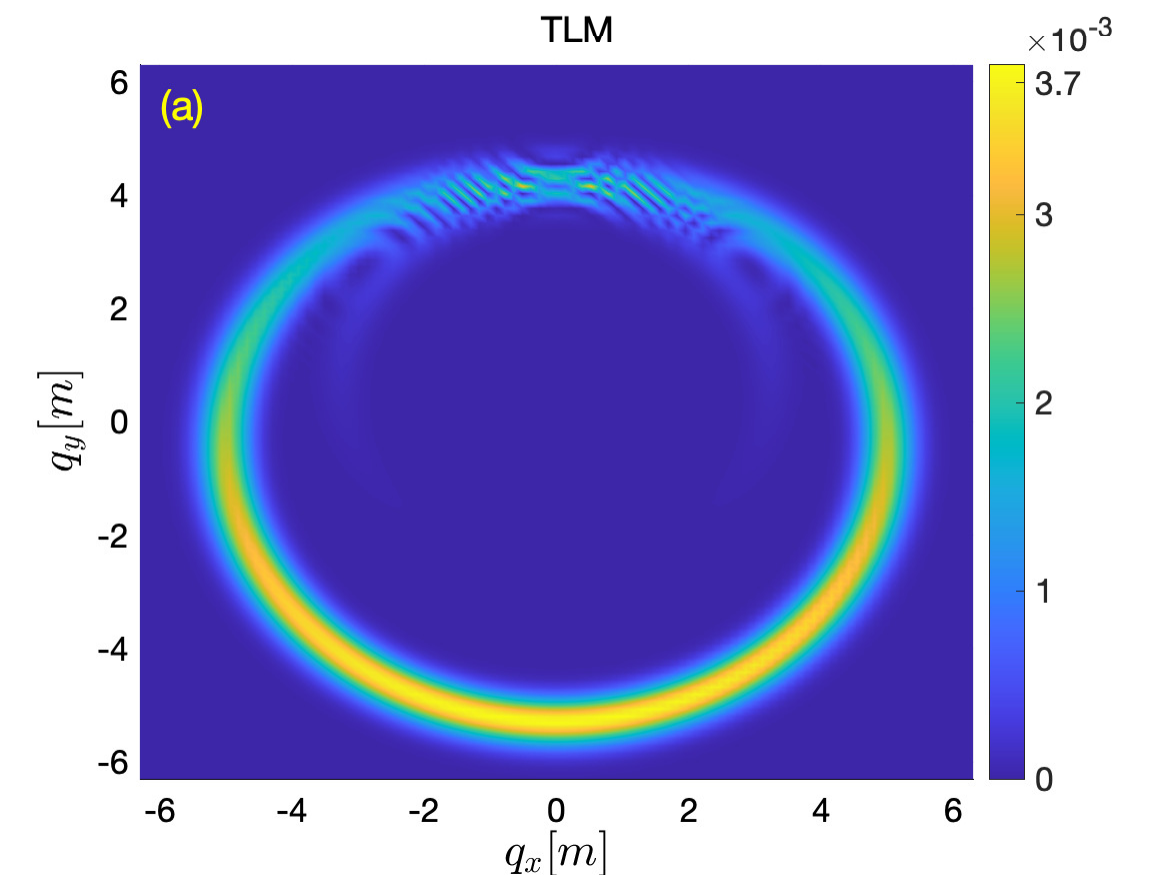}
\includegraphics[width=0.49\textwidth]{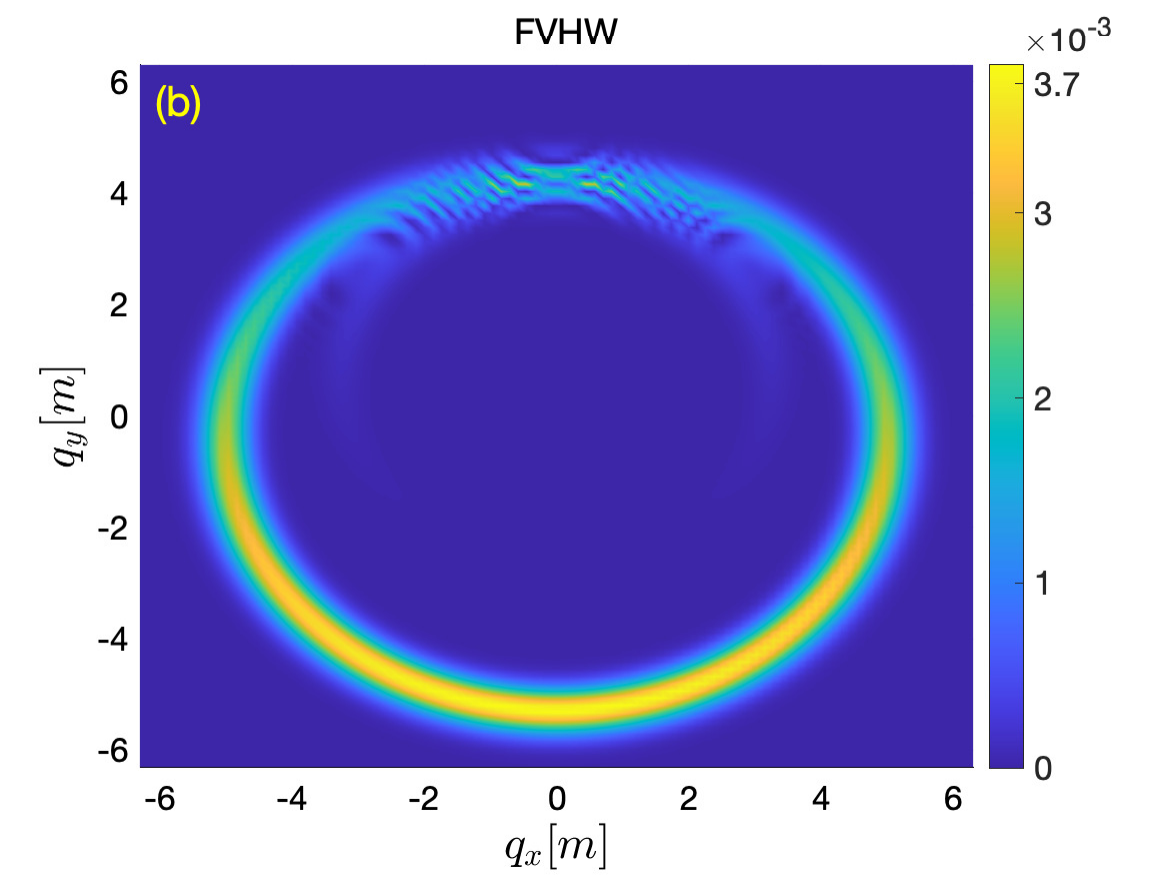}
\caption{Electron momentum spectra under the CP electric field (where $q_z=0$). (a) and (b) are the numerical results of scalar TLM and FVHW methods. Other field parameters are $\varepsilon_{0}=0.5$, $N=5$ and $\omega_{0}=0.1 m$.}
\label{fig:11}
\end{figure}
The momentum spectra of the scalar TLM and FVHW methods are shown in Figs.~\ref{fig:11}(a) and (b).
One can mainly find one concentric ring. This demonstrates that the effect mainly comes from the tunneling dominated process, because Keldysh parameter $\gamma_{\omega}=0.2$.
Moreover, we can also see that the radius of the ring in the momentum spectra in Figs.~\ref{fig:11} are larger than that of the radius of the ring in the momentum spectra in multiphoton dominated process in Fig.~\ref{fig:10}.
It is caused by the tunneling dominated process~\cite{Amat:2023vwv}.
Hence, the peak momentum of the momentum spectra in Figs.~\ref{fig:11}(a) and (b) are far away from the center, and it leads to the appearance of large rings in the momentum spectra.

We further find that the two results from the scalar TLM and FVHW methods are almost the same, i.e., the ratio of the two momentum spectra $R_B$ is about $1$. One can clearly see the interference effect in the momentum spectra. This indicates that we can observe the interference effect on the momentum spectra obtained from scalar TLM. The scalar TLM is completely equivalent to the quantum mechanical scattering problem. Thus, the origin of the interference effect is the resonances in the scattering problem as the shape of the potential $-\omega^2_{\mathbf{q}}(t)$ changes with the momentum $\mathbf{q}$ \cite{Hebenstreit:2009km}.

In short, the two methods are consistent for the tunneling dominated process.

\section{Dirac-Heisenberg-Wigner formalism}\label{ap:d}

The starting point is the Dirac equation with:
\begin{equation}\label{E1}
\left( i\gamma^{\mu}\partial_\mu- e \gamma^{\mu} A_\mu(\mathbf{x},t)-m\right)\Psi(\mathbf{x},t)=0,
\end{equation}
where $\partial_\mu=(\partial_t, \bm{\nabla})$, $e$ and $m$ are the electron charge and mass, respectively, $A_\mu(\mathbf{x},t)=(\varphi, -\mathbf{A})$ is the four-potential of electromagnetic fields.

sEq.~\ref{E1} can also be written as
\begin{equation}\label{E2}
i\frac{\partial}{\partial t}\Psi(\mathbf{x},t)=\mathcal{H}(\mathbf{x},t)\Psi(\mathbf{x},t),
\end{equation}
where the time-dependent Hamiltonian is
\begin{equation}\label{E3}
\mathcal{H}(\mathbf{x},t)=\bm{\alpha}\cdot\big[\mathbf{P}-e \mathbf{A}(\mathbf{x},t)\big]+\beta mc^2+q\varphi(\mathbf{x},t),
\end{equation}
where $\mathbf{P}=-i\bm{\nabla}$ is the canonical momentum operator, $\bm{\alpha}=\gamma^0\bm{\gamma}=\left(
           \begin{array}{cc}
             0 & \bm{\sigma} \\
             \bm{\sigma} & 0 \\
           \end{array}
         \right)$ and $\beta=\gamma^0$.

 The equal-time density operator for spinor QED, which is an equal-time commutator constructed by two field operators in the Heisenberg picture, can be expressed as \cite{Hebenstreit:2011}
\begin{align}\label{E4}
\hat{\mathcal{C}}_{spinor}(\mathbf{x}_1,\mathbf{x}_2,t)\equiv \exp\Big(-ie\int_{\mathbf{x}_2}^{\mathbf{x}_1} d\mathbf{x}'\cdot\mathbf{A}(\mathbf{x}',t)\Big) \left[\bar{\Psi} (\mathbf{x}_1,t),{\Psi}(\mathbf{x}_2,t) \right].
\end{align}
For the center of mass coordinate $\mathbf{x}=(\mathbf{x}_1+\mathbf{x}_2)/2$ and the relative coordinate $\mathbf{r}=\mathbf{x}_1-\mathbf{x}_2$, the density operator becomes
\begin{align}\label{E5}
\hat{\mathcal{C}}_{spinor}(\mathbf{x},\mathbf{r},t)=\exp\Big(-ie\int_{-1/2}^{1/2}d\lambda\,\mathbf{r}\cdot\mathbf{A}(\mathbf{x}+\lambda\mathbf{r},t)\Big) \left[\bar{\Psi} \Big(\mathbf{x}+\frac{\mathbf{r}}{2},t\Big),\Psi \Big(\mathbf{x}-\frac{\mathbf{r}}{2},t\Big)\right].
\end{align}

The Wigner function is defined as the vacuum expectation value of the Wigner operator which is the Fourier transform of the equal-time commutator of two Dirac field operators in the Heisenberg picture $\mathcal{C}_{spinor}(\mathbf{x},\mathbf{r},t)$ with respect to the relative coordinate
$\mathbf{r}$:
\begin{eqnarray}\label{E6}
\mathcal{W}_{spinor}(\mathbf{x},\mathbf{p},t)=\frac{1}{2}\int{d^3 \mathbf{r}\,
\langle\mathrm{vac}|\hat{\mathcal{C}}_{spinor}(\mathbf{x},\mathbf{r},t)|\mathrm{vac}\rangle e^{-i\mathbf{p}\cdot\mathbf{r}}}.
\end{eqnarray}
The second exponential function in the integrand on the right hand side of Eq. (\ref{E6}) is called the Wilson line factor which is introduced to keep gauge invariance. Furthermore, a straight line is chosen as the integration path in this factor to introduce a well defined kinetic momentum $\mathbf{p}$. Note that a Hartree approximation is used here, so the electromagnetic field is treated as a C-number field instead of a Q-number one.

Taking the time derivative of Eq.~(\ref{E6}) and applying the Dirac equation Eq. (\ref{E2}) with $\mathbf{A}(t)$ replaced by $\mathbf{A}(\mathbf{x},t)$, we can obtain the equation of motion for the Wigner function \cite{Hebenstreit:2011}:
\begin{equation}\label{E7}
  D_t\mathcal{W}_{spinor}=-\frac{1}{2}\mathbf{D}\cdot\left[\gamma^0\bm{\gamma},\mathcal{W}_{spinor}\right]
  \!-i\mathbf{\Pi}\cdot\left\{\gamma^0\bm{\gamma},\mathcal{W}_{spinor}\right\}
  \!-im\left[\gamma^0,\mathcal{W}_{spinor}\right]
  ,
\end{equation}
where $D_t$, $\mathbf{D}$ and $\mathbf{\Pi}$ denote the pseudo-differential operators
\begin{alignat}{6}
  &D_t&\ =\ &\ \ \partial_t \ &\ + \ & e  \int_{-1/2}^{1/2}{d\xi\,\mathbf{E}(\mathbf{x}+i\xi\partial_\mathbf{p}},t)\cdot\partial_\mathbf{p} \ , \nonumber
  \\
  \label{E8}
  &\mathbf{D}&\ =\ &\ \bm{\nabla} \ & + \ & e  \int_{-1/2}^{1/2}{d\xi\,\mathbf{B}(\mathbf{x}+i\xi\partial_\mathbf{p},t)
  \times\partial_\mathbf{p}} \ ,
  \\
  &\mathbf{\Pi}&\ =\ &\ \ \mathbf{p}\ &\ - \ & ie  \int_{-1/2}^{1/2}{d\xi\,\xi\,\mathbf{B}(\mathbf{x}+i\xi\partial_\mathbf{p},t)
  \times\partial_\mathbf{p}} \ . \nonumber
\end{alignat}

The Wigner function $\mathcal{W}_{spinor}(\mathbf{x},\mathbf{p},t)$ can be expanded in terms of a complete basis set $\{\mathbbm{1},\gamma_5,\gamma^\mu$,
$\gamma^\mu\gamma_5,\sigma^{\mu\nu}=:\frac{i}{2}[\gamma^\mu,\gamma^\nu]\}$ and $16$ irreducible components (DHW functions), scalar $\mathbbm{s}(\mathbf{x},\mathbf{p},t)$ (mass density), pseudoscalar $\mathbbm{p}(\mathbf{x},\mathbf{p},t)$ (pseudoscalar condensate density), vector $\mathbbm{v}_\mu(\mathbf{x},\mathbf{p},t)$ (${\mathbbm v}_0$ and ${\vec {\mathbbm v}}$ are the charge and current density respectively), axialvector $\mathbbm{a}_\mu(\mathbf{x},\mathbf{p},t)$ (polarization or spin current density) and tensor $\mathbbm{t}_{\mu\nu}(\mathbf{x},\mathbf{p},t)$ ($\mathbbm{t}_{0i}$ or $\mathbbm{t}_{j0}$ are the electric dipole-moment, and $\mathbbm{t}_{ij}$ is the magnetic dipole-moment, $i,j=1,2,3.$) \cite{Sheng:2019ujr}
\begin{equation}\label{E9}
  \mathcal{W}_{spinor}(\mathbf{x},\mathbf{p},t)=\frac{1}{4}\left(\mathbbm{1}\mathbbm{s}+i\gamma_5\mathbbm{p}
+\gamma^\mu\mathbbm{v}_\mu+\gamma^\mu\gamma_5\mathbbm{a}_\mu+\sigma^{\mu\nu}\mathbbm{t}_{\mu\nu}\right).
  \
\end{equation}

Correspondingly, one chooses the vacuum initial conditions as starting values. The non-vanishing values are
\begin{equation}\label{E10}
{\mathbbm s}_{\rm vac} = \frac{-2m}{\sqrt{{\mathbf p}^2+m^2}} \, ,
\quad  {\mathbbm v}_{i,{\rm vac}} = \frac{-2{ p_i} }{\sqrt{{\mathbf p}^2+m^2}}, \,
\end{equation}
where the subscript `$\rm vac$' represents the vacuum initial condition, $i=1,2,3$ represents the $x$, $y$ and $z$ directions.

In general, the equations of motions for the Wigner coefficients are integro-differential equations. Their numerical solution is featured by the challenging non-local nature of the respective pseudo-differential operators, see, {\it e.g.}, \cite{Hebenstreit:2011,Hebenstreit:2011wk,Kohlfurst:2015zxi}.

For the homogeneous electric field, these equations can be reduced to ordinary differential equations \cite{Blinne:2015zpa}. Here note that at most ten out of the sixteen Wigner coefficients are non-vanishing:
\begin{equation}\label{E11}
{\mathbbm w} = ( {\mathbbm s},{\mathbbm v}_i,{\mathbbm a}_i,{\mathbbm t}_i)
\, , \quad  {\mathbbm t}_i := {\mathbbm t}_{0i} -   {\mathbbm t}_{i0}  \, .
\end{equation}
Additionally, the kinetic momentum ${\mathbf p} $ is related to the canonical momentum
${\mathbf q}$ via
\begin{equation}\label{E12}
{\mathbf p}(t) = {\mathbf q} - e {\mathbf A} (t),
\end{equation}
which thus is time-dependent. In a next step one expresses the scalar Wigner coefficient by the one-electron distribution function $f({\mathbf q},t)$. The latter is related to the phase-space energy density,
\begin{equation}\label{E13}
\varepsilon = m {\mathbbm s} + p_i {\mathbbm v}_i.
\end{equation}
We can obtain the momentum distribution as
\begin{equation}\label{E14}
f({\mathbf q},t) = \frac 1 {2 \omega(\mathbf{q},t)} (\varepsilon - \varepsilon_{vac} ),
\end{equation}
where $\omega(\mathbf{q},t)= \sqrt{{\mathbf p}^2(t)+m^2}=
\sqrt{m^{2}+(\mathbf{q}-e\mathbf{A}(t))^{2}}$ is the electron's (or positron's) energy.

In addition it is helpful to redefine the following expressions \cite{Blinne:2015zpa,Blinne:2016yzv}
\begin{align}\label{E15}
&{\mathbbm s} (\mathbf{p}(t),t):= (1-f({\mathbf q},t)) {\mathbbm s}_{vac}-\mathbf{p}(t) \cdot v_i (\mathbf{q},t),\\  \label{E16}
&v_i (\mathbf{q},t):= {\mathbbm v}_i (\mathbf{p}(t),t) -(1-f({\mathbf q},t)) {\mathbbm v}_{i,vac} (\mathbf{p}(t),t),\\  \label{E17}
&a_i(\mathbf{q},t):=\mathbbm{a}_i(\mathbf{q},t),\\ \label{E18}
&t_i(\mathbf{q},t):=\mathbbm{t}_i(\mathbf{q},t).
\end{align}

The one-electron momentum distribution function $f(\mathbf{q},t)$ can be obtained by solving the following ten ordinary differential equations
\begin{equation}\label{E19}
\begin{array}{l}
\displaystyle
\dot{f}=\frac{e}{2\omega} \, \,  \mathbf{E}\cdot \mathbf{v},\\[2mm]
\displaystyle
\dot{\mathbf{v}}=\frac{2}{\omega^{3}}
\left( (e\mathbf{E}\cdot \mathbf{p})\mathbf{p}-e\omega^{2}\mathbf{E}\right) (f-1)
-\frac{(e\mathbf{E}\cdot \mathbf{v})\mathbf{p}}{\omega^{2}}
-2\mathbf{p}\times \mathbf{a} -2m \mathbf{t},\\[2mm]
\displaystyle
\dot{\mathbf{a}}=-2\mathbf{p}\times \mathbf{v},\\
\displaystyle
\dot{\mathbf{t}}=\frac{2}{m}[m^{2}\mathbf{v}+(\mathbf{p}\cdot \mathbf{v})\mathbf{p}].
\end{array}
\end{equation}
The initial condition values are selected as $f(\mathbf{p},-\infty)=\mathbf{v}(\mathbf{p},-\infty)= \mathbf{a}(\mathbf{p},-\infty)=\mathbf{t}(\mathbf{p},-\infty)=0$ in order to perform the calculation.  We can further obtain the number density of created pairs by integrating the distribution function $f(\mathbf{q},t)$ over all momenta at asymptotically late times $t\to +\infty$:
\begin{equation}\label{E20}
  n(t\rightarrow\infty)= \lim_{t\to +\infty}\int\frac{d^{3}q}{(2\pi)^ 3}f(\mathbf{q},t) \, .
\end{equation}

For time-dependent two-component fields $A^\mu(x)=(0,A_x(t),A_y(t),0)$, the spin-dependent\\
electron momentum distribution in the $z$ direction can be written as \cite{Blinne:2015zpa}
\begin{align}\label{E21}
f^{\bf s}&=\frac{1}{2} \left( f+{\bf s} \, \delta f_\mathrm{sc} \right),
\end{align}
where ${\bf s}$ denotes the spin polarization number, i.e., ${\bf s}=+1$ and ${\bf s}=-1$ for spin up and down. The semiclassical spin projection, chiral and magnetic momentum asymmetries are defined as \cite{Blinne:2015zpa}
\begin{align}\label{E22}
\delta f_\mathrm{sc}&=\frac{q_z}{\epsilon_\perp}\delta f_\mathrm{c}+\frac{m}{\epsilon_\perp}\delta f_{\mu_z},\\ \label{E23}
\delta f_\mathrm{c}&=\frac{1}{2\omega} \mathbf{p}\cdot\mathbf{a},\\ \label{E24}
\delta f_\mathrm{\mu_z}&=\frac{1}{2\omega}\left( m a_z+(\mathbf{p}\times\mathbf{t})_z \right),
\end{align}
where
\begin{align}\label{E25}
\epsilon_\perp=\sqrt{q_z^2+m^2}.
\end{align}

\end{document}